\documentclass[pra,twocolumn,noshowpacs,preprintnumbers,amsmath,amssymb,superscriptaddress]{revtex4}

\usepackage[latin1]{inputenc}
\usepackage[english]{babel}
\usepackage[dvips]{graphicx}
\usepackage{amsmath,amsbsy,amsfonts,amssymb}
\usepackage{graphicx}
\usepackage{dcolumn}
\usepackage{bm}

\begin{document}
\title{\textbf{Diffraction effects on light-atomic ensemble quantum interface}}
\author{J.~H.~M\"{u}ller}
\email{muller@nbi.dk}
\author{P.~Petrov}
\affiliation{QUANTOP, Niels Bohr Institute, Copenhagen University,
Blegdamsvej 17, DK-2100 København, Denmark}
\author{D.~Oblak}
\affiliation{QUANTOP, Niels Bohr Institute, Copenhagen University,
Blegdamsvej 17, DK-2100 København, Denmark} \affiliation{Institute
of Physics and Astronomy, University of Aarhus, Ny Munkegade
Bldg.520, DK-8000 Aarhus, Denmark}
\author{C.~L.~\surname{Garrido Alzar}}
\affiliation{QUANTOP, Niels Bohr Institute, Copenhagen University,
Blegdamsvej 17, DK-2100 København, Denmark}

\author{S.~R.~de Echaniz}
\affiliation{IFCO-Institut de Ci\`{e}ncies Fot\`{o}niques, Jordi
Girona 29, Nexus II, E-08034 Barcelona, Spain}
\author{E.~S.~Polzik}
\homepage{http://quantop.nbi.dk/index.html}
 \affiliation{QUANTOP, Niels
Bohr Institute, Copenhagen University, Blegdamsvej 17, DK-2100
København, Denmark}

\date{\today}

\begin{abstract}
We present a simple method to include the effects of diffraction
into the description of a light-atomic ensemble quantum interface
in the context of collective variables. Carrying out a scattering
calculation we single out the purely geometrical effect and apply
our method to the experimental relevant case of Gaussian shaped
atomic samples stored in single beam optical dipole traps probed
by a Gaussian beam. We derive simple scaling relations for the
effect of the interaction geometry and compare our findings to the
results from one dimensional models of light propagation.

\end{abstract}

\pacs{32.80Pj, 42.50}

\maketitle
\section {Introduction}
Coupling collective variables of atomic ensembles to propagating
modes of the electromagnetic field has been identified as an
efficient tool to engineer the states of atoms and light at the
quantum level. Several proposals for spin-squeezing, mapping of
quantum states between light and atoms, i.e. quantum memory
operations, creation of macroscopic entanglement, and
teleportation of atomic states have been published \cite{KuMoPo,
KuzmichEuro, duan, Kuzmich2000, Kozhekin, wiseman, Kraus,
Fiurasek}. A number of these proposals have actually been verified
experimentally using atomic samples stored in vapor cells as well
as in magneto optical traps \cite{Hald, kuzmichprl, briannature,
Schori}. The efficiency of the coupling is often discussed
resorting to effective 1-D models for the light propagating
through a homogeneous sample and the optical depth is found to be
the essential parameter determining the coupling strength. This
naturally suggests the use of cold and trapped atomic samples,
where long coherence times for collective variables are possible,
with a shape mimicking a 1-D string of atoms to maximize column
density for a given number of atoms. However, while the effective
1-D models work well for wide and nearly homogeneous samples the
question of light diffraction from small, inherently inhomogeneous
samples and its impact on coupling efficiency immediately arises
when cold and trapped atomic samples are used \cite{LuMingDuan}.
In a recent work \cite{Kuzmich04} aspects of spatial inhomogeneity
have been addressed via the introduction of new asymmetric
collective variables.

The purpose of the present paper is twofold. We present a simple
and effective semi-analytic method to identify the spatial light
mode the atomic sample couples to and use this method to find
optimum shapes for atomic samples. In addition we will derive
simple scaling parameters describing the effect of the shape of
the atomic sample which allow us to make a link to the established
1-D models and quantify the effect of diffraction on the coupling
strength. Albeit we are ultimately interested in measuring and
manipulating quantum states, we find it instructive to solve the
underlying classical scattering problem first to single out the
purely geometric effects, and we try to avoid hiding useful
practical information behind the much more intricate mathematical
formalism needed to solve the full quantum problem. Our approach
delivers valuable information for designing experimental
configurations, provides intuitive pictures, and may serve as a
starting point for a more elaborate quantum theoretical treatment.

The remainder of the paper is organized as follows. In Section 2
we review briefly the quantum description of two modes of light
coupled coherently to atoms in terms of collective variables and
introduce typical experimental configurations. Using this
framework we motivate the use of the coherently scattered power by
the atomic sample as the relevant parameter to optimize for the
quantum coupling. Section 3 presents our model for calculating the
scattering mode and the scattering efficiency of the atomic sample
for different geometries together with some remarks on the
assumptions made and the range of validity of the model. In
Section 4 we apply this model to atoms stored in a single beam
optical dipole trap. In Section 5 we provide a link between our
classical calculation and the results from effective 1-D models.
Section 6 summarizes our results and points out possible
extensions of our model.
\section{Collective light-atom coupling}
\label{collective}
\begin{figure*}
  \centering
  \includegraphics[width=0.65\textwidth]{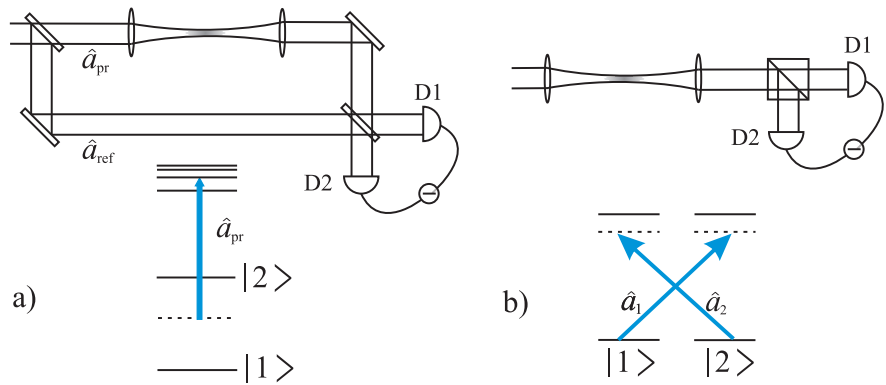}
  \caption{(color online) Mach-Zehnder and polarization interferometry setups for quantum coupling of collective variables
  of cold atomic samples to light.}\label{fig:interferometer}
\end{figure*}

 The hermitian part of the interaction of a
two-mode pulse of off-resonant light with an ensemble of atoms
with two ground states residing in a container with
cross-sectional area $A$ and length $L$ can be described by a
Hamiltonian of the form (see e.g.\cite{Happer, Kuzmich1, Kraus,
Footnote1}):
\begin{equation}\label{Hamilton}
\hat{H_{c}} =  \hbar \frac{2\sigma_{0}}{A}\left(
\frac{\Gamma}{2\Delta}\right) \int_{0}^{L}
\hat{S}_{z}(z,t)\hat{J}_{z}(z,t) dz
\end{equation}
The factor in front of the integral describing the strength of the
coupling of the collective variables contains the single-atom
reactive response to radiation characterized by the cross-section
$\sigma_{0}$, the detuning $\Delta$ in units of the natural
linewidth $\Gamma$, and the cross-sectional area of the light mode
governing the electric field strength per photon occupying the
mode. In eq.(\ref{Hamilton}) collective operators for light
$\hat{\mathbf{S}}$  and atoms $\hat{\mathbf{J}}$ are introduced.
For light we can write $\hat{\mathbf{S}}$ in terms of continuous
mode creation and annihilation operators with boson commutation
rules:
$\hat{S}_{x} = \frac{1}{2}
(\hat{a}_{2}^{+}\hat{a}_{1}+\hat{a}_{1}^{+}\hat{a}_{2}),
\hat{S}_{y} = \frac{1}{2i}
(\hat{a}_{2}^{+}\hat{a}_{1}-\hat{a}_{1}^{+}\hat{a}_{2}),
\hat{S}_{z} = \frac{1}{2}
(\hat{a}_{2}^{+}\hat{a}_{2}-\hat{a}_{1}^{+}\hat{a}_{1})$.
For simplicity of notation the time and spatial dependence of the
operators has been suppressed. The mode index can specify for
instance two fields propagating in the two legs of a Mach-Zehnder
interferometer (Fig.\ref{fig:interferometer}a) or left and right
circular polarized modes of a single laser beam
(Fig.\ref{fig:interferometer}b), in which case the operator
expressions describe the instantaneous Stokes vector of the light
\cite{Footnote2}.
By construction the collective operators follow the usual
commutator algebra for vector operators. For atoms, restricting us
for simplicity to the case of two ground-states only, we define in
analogy:
$\hat{J}_{x} = \sum_{j=1}^{N_{at}(z)} \frac{1}{2} (|2\rangle
_{j}\langle 1| + |1\rangle _{j}\langle 2|), \hat{J}_{y} =
\sum_{j=1}^{N_{at}(z)} \frac{1}{2i}(|2\rangle _{j}\langle 1| -
|1\rangle _{j}\langle 2|),  \hat{J}_{z} = \sum_{j=1}^{N_{at}(z)}
\frac{1}{2} (|2\rangle _{j}\langle 2| - |1\rangle _{j}\langle
1|)$,
where the sum extends over the number of atoms $N_{at}(z)$
residing in a thin slice of the interaction volume at position $z$
\cite{Footnote3}. A continuum limit is taken by shrinking the
thickness $\delta z$ of the slices to zero. The three operators
$\hat{J}_{i}$ are the macroscopic Bloch-vector components for the
2-level system formed by the two ground states. Inherent in this
description are the assumptions that the transverse mode function
is a frozen variable, hence the 1-D integration over the Hamilton
density in eq.(\ref{Hamilton}), and that the atoms scatter light
coherently only into the input modes. This is a good approximation
for atomic ensembles and light fields transversally much wider
than an optical wavelength and not too long samples. In addition,
since the operator nature of atomic position is suppressed, the
spatial density distribution of the atoms is not changed by the
interaction with the radiation in this model.

The dynamics generated by eq.(\ref{Hamilton}) at the level of
expectation values describes differential phase shifts of the
light modes due to coherent forward scattering by the atoms in the
two ground states and differential shifts of the atomic energy
levels due to the intensity difference of the two light modes.
This dynamics can be conveniently visualized with the help of
coupled Bloch spheres for light and atom variables. Of particular
interest for practical quantum state engineering is the case of
initial conditions with $\hat{J}_{x}^{in}$ and $\hat{S}_{x}^{in}$
large and classical, i.e. uncorrelated coherent states with
expectation values on the order of half the total number of atoms
and photons respectively. In this case the dynamical evolution of
the system is driven entirely by the fluctuations of light and
atomic variables. In the experimental setup sketched in
Fig.\ref{fig:interferometer}b this corresponds to linearly
polarized input light and all atoms in a balanced coherent
superposition of the two ground states, such that there is a
macroscopic magnetic polarization in the plane orthogonal to the
direction of propagation. Linearizing the spectrum of the
operators around their expectation values and integrating the
dynamics, to leading order, input/output relations can be written
for the fluctuations of the collective operators orthogonal to the
coherent excitation \cite{Footnote4, Sinatra}:
\begin{eqnarray}
\delta\hat{J}_{y}^{out} & = & \delta\hat{J}_{y}^{in} - \frac{2
\sigma_{0}}{A}
\frac{\Gamma}{2\Delta} <\hat{J}_{x}> \delta\hat{S}_{z}^{in}\nonumber\\
\delta\hat{J}_{z}^{out}& = & \delta\hat{J}_{z}^{in} \nonumber\\
\delta\hat{S}_{y}^{out} & = & \delta\hat{S}_{y}^{in} - \frac{2
\sigma_{0}}{A}
\frac{\Gamma}{2\Delta} <\hat{S}_{x}>  \delta\hat{J}_{z}^{in}\nonumber\\
\delta\hat{S}_{z}^{out} & = & \delta\hat{S}_{z}^{in} \nonumber
\end{eqnarray}

From these relations we can calculate the variance of a set of
measurements on identically prepared systems. A set of measurement
records of $\hat{S}_{y}^{out}$ done over some time interval  will
show a variance:
\begin{equation}\label{Variance}\nonumber
Var(\hat{S}_{y}^{out}) = \frac{n_{ph}}{4} + \left(\frac{2
\sigma_{0}}{A}\frac{\Gamma}{2\Delta}
\frac{n_{ph}}{2}\right)^{2}\frac{N_{a}}{4}
\end{equation}
Here the first term reflects the noise of the coherent input state
of light, with $n_{ph}$ the total number of probe photons, while
the second term describes the excess fluctuations imprinted onto
the light by the fluctuations of the atomic variable and amplified
by the large coherent amplitude of the conjugate light variable.
From the measurement record we can infer a value for the atomic
variable with a confidence interval limited by the first term,
i.e. with a signal to noise ratio of:
\begin{equation}\label{SignaltoNoise1-D}
 \left(\frac{S}{N}\right)^{2}_{1-D} =
 \frac{\sigma_{0}^{2}}{A^{2}}N_{at}n_{ph}\left(\frac{\Gamma}{2\Delta}\right)^{2}
\end{equation}
Projection of the atomic state by the destructive measurement on
light, will leave the collective atomic variable in a state with
fluctuations reduced below the standard quantum limit if the
signal to noise ratio is finite. In practical terms this means
that repeated measurements on the same atomic system will show a
covariance below the projection noise \cite{Wineland}. To the
extent the measurement is nondestructive, i.e. after proper
inclusion of spontaneous emission \cite{Hammerer}, this fact can
be exploited to write, store, and retrieve quantum information to
and from the atomic variable. Instead of measuring destructively
the light pulse after a single pass through the atomic ensemble
one can allow for multiple interaction together with appropriate
switching of the coupling Hamiltonian between consecutive passes.
This way unconditional quantum state preparation is possible and
the measurement serves merely as a verification of successful
state preparation \cite{Hammerer, Braunstein, Fiurasek}. In this
more general setting the above defined signal to noise ratio plays
the role of a coupling strength $\kappa^2$:
\begin{eqnarray}\label{couplingstrength}
 \kappa^2 & = &
 \frac{\sigma_{0}^{2}}{A^{2}}N_{at}n_{ph}\left(\frac{\Gamma}{2\Delta}\right)^{2}
 \\ \nonumber & = & \alpha_{0} \eta
\end{eqnarray}
which can be conveniently expressed as the product of integrated
spontaneous emission rate
${\eta}~=~{n_{ph}}(\sigma_{0}/A)~({\Gamma}/2\Delta)^{2}$ and the
optical depth or column density $\alpha_{0} = N_{at}
(\sigma_{0}/A)$, explaining the use of the optical depth as a
figure of merit for collective variable light-atom coupling
\cite{Braunstein}.
 The interaction geometry, of course, stays the
same, so diffraction of light discussed in the following has the
same impact on the coupling strength in these schemes.

After this brief review of the 1-D quantum model, we look again at
the experimental configuration in Fig.\ref{fig:interferometer}a
\cite{Oblak}, seeking this time a description in purely classical
terms and pointing out where we go beyond the 1-D model. A cloud
of cold atoms residing in one arm of the interferometer is loaded
into a far-off-resonant dipole trap created by a focused gaussian
beam. Probe light with a wave-vector $k = 2\pi/\lambda$ enters the
interferometer from the left and passes through the atomic sample.
The light experiences a phase shift caused by the refractive index
of the sample, which is determined by the atomic density and the
distribution of population among the two different sub-levels.
Given an atomic density distribution the light carries thus
information about the level populations after passage through the
sample.  In turn, the atomic energy levels are shifted
differentially during the interaction and thus information about
the light is deposited in the coherence between the atomic levels.
To assess the sensitivity of a measurement of the atomic
population difference we look at the signal by the balanced
homodyne detector (with quantum efficiency $\varepsilon$) on the
right upon detection of a pulse of light (frequency $\omega$) of
duration $\tau$. The detector signal $S_{D}$, i.e. the
photocurrent $i_{s}$ in units of the elementary charge $e$
integrated over the pulse duration, in the presence of atoms is
given by:
\begin{eqnarray} \label{signal}
S_{D} & = & \int_{0}^{\tau}\frac{i_{s}}{e}dt \nonumber \\
& =  & \varepsilon\frac{\tau}{\hbar \omega}
\frac{c\varepsilon_{0}}{2}\bigg [
2|E_{sc}||E_{ref}|\int_{A_{D}}M_{sc}M_{ref}dA \nonumber \\  & +
 & 2|E_{probe}||E_{ref}|  \int_{A_{D}}M_{probe}M_{ref}dA \bigg ]
\end{eqnarray}
Here $E_{probe} (E_{ref})$ denotes the initial field in the probe
(reference) arm. The integrals over the detector area describe the
overlap of the transverse mode  functions $M_{sc}, M_{probe},
M_{ref}$ in the detector plane and contain also the oscillatory
dependence on the path length difference of the interferometer. We
separate explicitly the field scattered by the atoms $E_{sc}$ from
the probe field $E_{probe}$, since in general it will not be in
the same spatial mode and have a different phase. We can adjust
the path length difference such to make the second integral in
eq.(\ref{signal}) vanish - this corresponds to the preparation of
a large $\langle\hat{S}_{x}\rangle$ and a measurement of
$\langle\hat{S}_{y}\rangle$ with the detector in the language of
the quantum description. We note that for dispersive scattering in
the far field, probe wave and scattered wave are nearly $\pi /2$
out of phase, so the first term in eq.(\ref{signal}) will take its
maximum value whenever the second term vanishes. The detector
signal stems then from the interference of the reference wave with
the scattered wave only. The atomic contribution and consequently
the sensitivity of the measurement will thus be a monotonic
function of both the coherently scattered power by the atoms and
of the geometric overlap of the wavefronts of scattered wave and
reference wave. Carrying out the overlap integral assuming perfect
matching of the reference wave to the scattered wave, we write for
the detector signal power $S^{2}$:
\begin{eqnarray}\label{eq:signalphoto}
 S^{2}=2\varepsilon^{2}\left(\frac{\tau}
 {\hbar \omega}\right)^{2}P_{sc}P_{ref}\nonumber
\end{eqnarray}
The noise will be limited from below by the shot noise of the
detected photons \cite{Footnote5}. We write the corresponding
noise power $N^{2}$ as:
\begin{eqnarray}\label{eq:noisephoto}
 N^{2}=\frac{\varepsilon\tau}{\hbar\omega} 2 P_{ref}\nonumber
\end{eqnarray}
Since we want to assess the sensitivity with reference to the
fluctuations of the population difference in an uncorrelated
atomic sample we have to normalize our signal power to this
variance, i.e. to a half of the total number of atoms. With this
we arrive at an equivalent expression for the signal to noise
ratio (SNR) or coupling strength as:
\begin{equation}\label{eq:SNR+at_fluct}
   \kappa^2 = \left(\frac{S}{N}\right)^{2}=
   \frac{2\varepsilon\tau}{\hbar\omega}\frac{P_{sc}}{N_{at}}
  = 2\varepsilon \frac{n_{ph}^{sc}}{N_{at}}
\end{equation}

Here the coupling strength is expressed in terms of the -
admittedly artificial quantity (see section \ref{relation}) -
total scattered power per atom. We conclude that optimizing the
sensitivity is equivalent to maximizing the scattering efficiency
and adapting the reference wavefront to the scattered wave.
Trivially the scattered power can be increased by increasing the
input power to the interferometer. But here upper limits are set
by saturation of the detectors and the requirement for low
spontaneous emission rate. We choose therefore to perform the
search for optimal shapes of the atomic samples and of the probe
beam in the next section under conditions of constant probe power.
Given a maximum pulse energy the detectors can withstand, the
integrated spontaneous emission rate can be set to a desired value
by choosing the single atom scattering cross section via the
detuning.
 In the next section we calculate the
scattering efficiency and the scattered wavefront for selected
geometries.
\section{Scattering model}
\label{Scattering model}
In order to calculate the stationary
diffracted field by the atomic density distribution we model the
sample as an ensemble of fixed (i.e. infinitely heavy) point
scatterers and make use of a Born approximation \cite{deVries,
BornWolf}. Assuming fixed positions we neglect Doppler and recoil
shifts. Since we expect coherent diffraction to occur mainly close
to the forward direction and we are interested in samples at
ultralow temperatures, Doppler shifts will play a negligible role.
On time scales short compared to the recoil time - the time needed
for an atom to travel a distance of an optical wavelength at one
recoil velocity - we can assume that spatial correlations are
frozen, i.e. neglect of the recoil to the atom during scattering
is valid \cite{Footnote6}. In the first Born approximation we
neglect multiple scattering events and calculate the total
scattered field as the sum  of the fields scattered by independent
single scatterers out of the probe field. This approach is
justified as long as the contribution from all other scatterers to
the local field seen by an individual scatterer is small compared
to the probe field which demands e.g. negligible absorption. This
condition is also helped by destructive interference of scattered
waves for the case of balanced sub-level populations in the
experimental configuration introduced in the previous section. The
condition of independent scatterers can only be met for not too
high atomic density ($n_{at}< k^{3}$) otherwise resonant
dipole-dipole interaction can change the single atom scattering
properties appreciably.

\subsection{Scattering integral}
As the first ingredient we need the scattering amplitude $f$ for a
single atom. In our calculations we replace the p-wave scattering
amplitude by an isotropic amplitude f equal in magnitude to the
true forward scattering amplitude. This assumption greatly
simplifies analytic calculations and will be justified a
posteriori by the observation that constructive interference
occurs only close to the forward direction. Integrating $|f|^{2}$
over the solid angle renders the scattering cross-section at a
given detuning $\Delta$ of the probe laser from the atomic
resonance \cite{Footnote8}. For the experimentally relevant case
of alkali atoms our choice for the scattering cross section is
valid for any sub-level of the ground state probed by linearly
polarized light provided the detuning $\Delta$ is large compared
to the excited state hyperfine splitting:
\begin{eqnarray}
 \int_{4\pi}|f|^{2}d\Omega=\sigma_{0}\frac{1}
 {1+\left(\frac{2\Delta}{\Gamma}\right)^{2}}\nonumber \\
 \sigma_{0}=\frac{\lambda^{2}}{2\pi}\nonumber \\
 f = -\lambda\sqrt{\frac{3}{8\pi^{2}}}
 \frac{1 + i \left(\frac{2\Delta}{\Gamma}\right)}
 {1+ \left(\frac{2\Delta}{\Gamma}\right)^2}
\nonumber
\end{eqnarray}
We remark here that this classical treatment includes the full
response of the atom to the incident field, i.e. also the coupling
to all empty modes of the electromagnetic field. There is no
distinction between spontaneous and induced emission and the only
assumption made is that the response of the atom is linear in the
incident field, i.e. the atomic transition is not saturated.
Naturally, since we treat the field as a classical variable, the
statistical properties of the field, e.g. also the frequency
spectrum of detected scattered light, are not described correctly.
However, in the experimentally relevant case of large detuning and
low intensity scattering is almost entirely elastic \cite{Mollow}
and in the first Born approximation these shortcomings do not
enter the problem, since exchange of photons between the atoms,
i.e. multiple scattering, is neglected.

\begin{figure}
  \centering
  \includegraphics[width=0.40\textwidth]{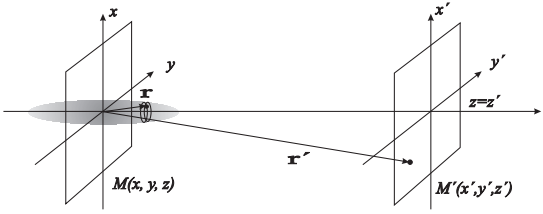}
  \caption{Huygens-Fresnel propagation used in the model}\label{fig:hf_integral}
\end{figure}

 The total scattered wave is the sum of the waves
scattered by individual scatterers $j$ ($j=1..N_{at}$) and can be
expressed in terms of electromagnetic field vector in complex
notation as [Fig.\ref{fig:hf_integral}]:
\begin{equation}
\vec{E}_{sc}(\vec{r'})=\sum^{N_{at}}_{j=1}\vec{E}_{probe}(\vec{r_{j}})f
\frac{\exp(-ik|\vec{r'}-\vec{r_{j}}|)}{|\vec{r'}-\vec{r_{j}}|}
\nonumber
\end{equation}

The atomic density or density of the point scatterers can be
written in the form:
\begin{equation}
 n(\vec{r})=\sum^{N_{at}}_{j=1}\delta(\vec{r}-\vec{r_{j}})
\nonumber
\end{equation}
with natural normalization condition giving the number of atoms in
the atomic sample:
\begin{equation}
 N_{at}=\int_{\mathbb{R}^{3}}dr^{3}n(\vec{r})
\nonumber
\end{equation}

In the following we use a continuous density distribution
according to a smooth probability distribution to find a particle
in a small volume element, again suitably normalized to the total
number of particles. This averaging procedure eliminates large
angle scattering off the microscopic density fluctuations, which
is equivalent to the single particle Rayleigh scattering
background for a discrete distribution of scatterers, i.e.
spontaneous emission at low saturation parameter \cite{deVries}.
Equally Bragg scattering of light from a regular distribution of
atoms on the wavelength scale is lost by the coarse graining. The
scattered field at some observation point $\vec{r'}$ outside the
sample in integral form becomes:
\begin{equation}\label{Escatt}
\vec{E}_{sc}(\vec{r'})=\int_{\mathbb{R}^{2}}dxdy\int^{L}_{-L}dz\vec{E}_{probe}
(\vec{r})n(\vec{r})f\frac{\exp(-ik|\vec{r'}-\vec{r}|)}{|\vec{r'}-\vec{r}|}
\end{equation}
In the paraxial domain (small angles to the optical axis), where
we expect constructive interference of scattering amplitudes to be
concentrated, we can approximate the spherical wave propagator in
eq.(\ref{Escatt}) by using a Fresnel expansion formula for the
distance $|\vec{r'}-\vec{r}|$

\begin{equation}
|\vec{r'}-\vec{r}|\simeq z'-z +
\frac{1}{2}\frac{x'^{2}+y'^{2}+x^{2}+y^{2}-2xx'-2yy'}{z'-z}
\nonumber
\end{equation}
in the phase factor, while we use $|\vec{r'}-\vec{r}|\simeq z'-z$
in the less critical denominator. Inserting this we can write the
propagator in eq.(\ref{Escatt}) as:
\begin{eqnarray}
K(|\vec{r'}-\vec{r}|)& \simeq &
\frac{\exp(-ik(z'-z))}{(z'-z)}\exp\left\{ik\frac{xx'+yy'}{z'-z}\right\}
\nonumber \\
& & \times\exp\left\{-ik\frac{x^{2}+y^{2}}{2(z'-z)}\right\}
\exp\left\{-ik\frac{x'^{2}+y'^{2}}{2(z'-z)}\right\} \nonumber
\end{eqnarray}
Since we want to describe free diffraction of probe light and
scattered light on an equal footing, we choose the incident probe
beam not as a plane wave but rather as gaussian with parameters
$w(z), R(z), \Phi(z)$ being the beam radius, wavefront radius and
Guoy phase, respectively:

\begin{eqnarray}
 & & \vec{E}_{probe}(x,y,z)  =  \vec{E}_{0}\frac{w(0)}{w(z)}
 \nonumber \\
& & \times
\exp\left\{-i[kz-\Phi(z)]-\frac{x^{2}+y^{2}}{w^{2}(z)}-ik\frac{x^{2}+y^{2}}{2R(z)}\right\}
\nonumber
\end{eqnarray}

As a realistic model for the density distribution of the trapped
sample we choose a Gaussian function. This corresponds to the
equilibrium shape of a thermal distribution of atoms residing
inside a harmonic oscillator potential. For low enough temperature
with respect to the trap depth this is a good description for a
dipole trapped sample. In particular, for the transverse
dimensions, where we carry out the the integration of the
scattering integral analytically, this choice simplifies the
mathematics. We note in passing that a generalization to arbitrary
transverse distributions by expansion into Hermite polynomials is
possible though analytically cumbersome.

The atomic density distribution in the transverse direction has a
radius $w_{a}$, which depends on $z$ due to the weaker confinement
by the dipole trap laser beam (wavelength $\lambda_{dip}$) away
from its minimal beam waist. In the longitudinal direction (along
the propagation axis of the probe beam) it is described by a
$1/e$-length parameter $L_{0}$.
\begin{eqnarray}
n(x,y,z) & = &
\frac{N_{at}}{\pi^{3/2}L_{0}w_{a}^{2}(z)}\exp\left[-\frac{x^{2}+y^{2}}{w_{a}^{2}(z)}-\frac{z^{2}}{L_{0}^{2}}\right]
\nonumber \\
w_{a}(z) & = & w_{a}\sqrt{1+\left(\frac{z}{z_{dip}}\right)^{2}}
\nonumber \\
z_{dip} & = & \frac{\pi w_{a}^{2}}{\lambda_{dip}}
\end{eqnarray}

Finally, the scattered wave field can be evaluated by solving the
integral
\begin{eqnarray} \label{E-scat}
\vec{E}_{sc}(\vec{r'})=\int_{\mathbb{R}^{2}}dxdy\int^{L}_{-L}dz
\vec{E}_{probe}(\vec{r})n(\vec{r})f K(|\vec{r'}-\vec{r}|)
\end{eqnarray}
Here the integration over z is to be taken only over the length
effectively occupied by the sample, but cannot be extended beyond
the observation plane. We evaluate the scattered field
distribution in some distant observation plane ($M'$ in
Fig.\ref{fig:hf_integral}) by carrying out the integration over
the transverse coordinates of the sample analytically and
integrating numerically over the length of the sample. Using
standard software on a desktop PC a scattered field profile can be
calculated in several seconds allowing for fast interactive
optimization of parameters. Not surprisingly for our model
assumptions and the choice of the density distribution, we find
the scattered mode profile to be very close to Gaussian in all of
the studied cases and we can extract parameters like width and
radius of curvature by fitting to the corresponding mode profile.
The scattering efficiency is evaluated by calculating the total
scattered power in the observation plane.

\subsection{Qualitative considerations}
Before presenting results of the numerical calculations some
qualitative considerations are at hand to train our intuition for
the results to be expected. First, the total scattered power is
strictly proportional to the square of the number of atoms for a
fixed geometry of the sample. This is a simple consequence of our
continuum approximation for the density distribution and may at
first sight seem disturbing, but is of course natural for coherent
scattering, where constructive interference of single scattering
amplitudes occurs in phase-matched directions. Secondly, far
enough away from the sample all scattered waves will interfere
constructively in the strict forward direction, so the on-axis
scattered intensity for a wide probe beam will be independent of
the exact geometry of the sample. This implies that scattering
efficiency is determined effectively by the opening angle of the
scattering cone around the forward direction. Simple scaling
arguments can be derived for this opening angle by looking at
Fraunhofer diffraction from transversally and longitudinally
extended samples.

\begin{figure}[t]
  \centering
  \includegraphics[width=0.33\textwidth]{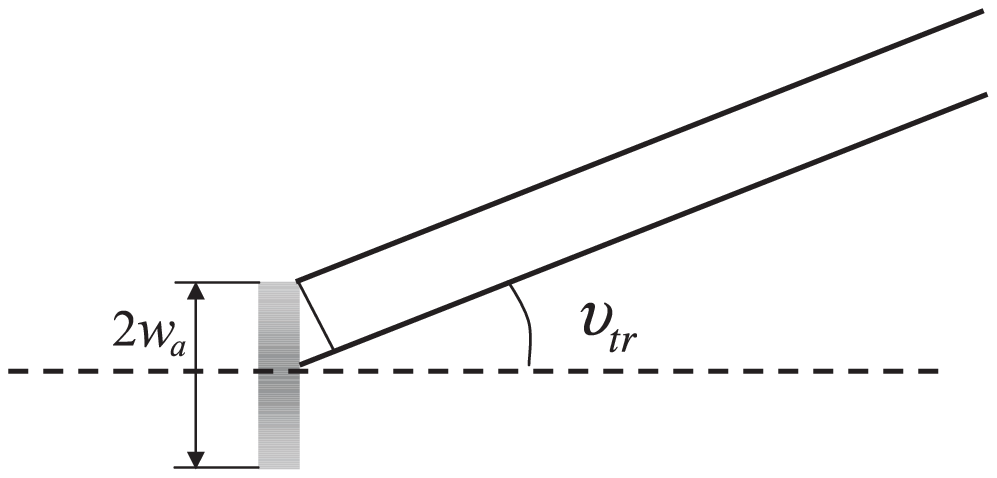}
  \includegraphics[width=0.43\textwidth]{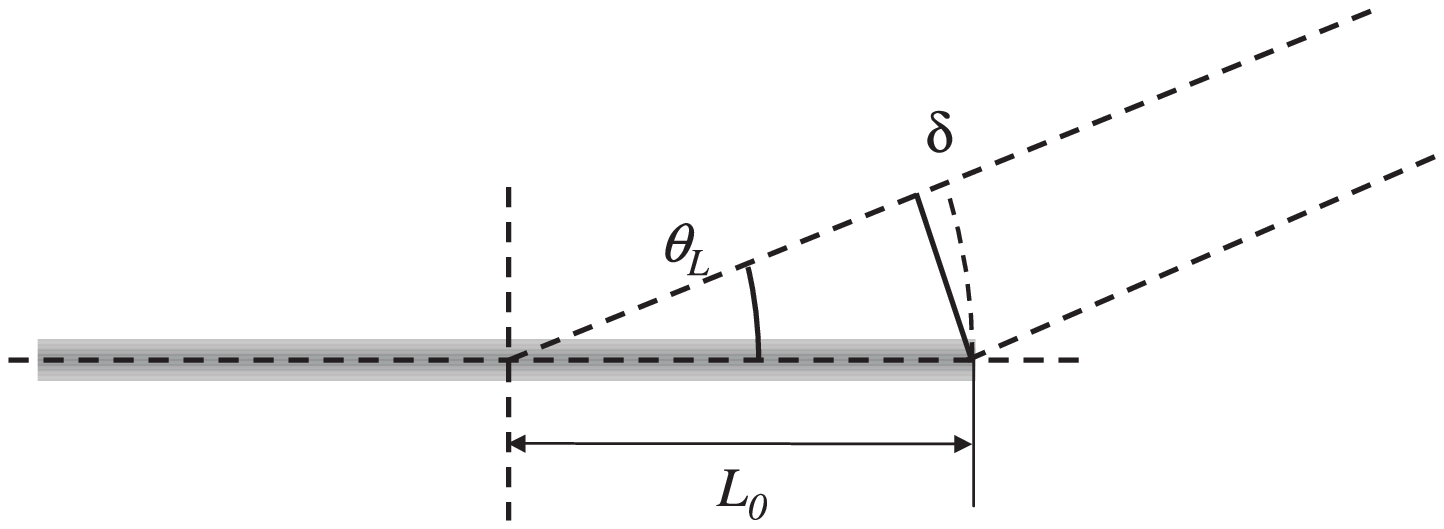}
  \caption{top: Diffraction limited scattering cone for Gaussian pancake-shaped sample; bottom:
  Diffraction limited scattering cone for Gaussian pencil-shaped sample}\label{fig:diff_cones}
\end{figure}

Let us first look at the diffraction cone of a short homogeneous
sample of width $2 w_{a}$ (Fig.\ref{fig:diff_cones}). To find the
angle where interference of scattered waves ceases to be
constructive we divide the sample in two halves. For a path length
difference of half a wavelength between the ends of a half,
destructive interference will occur, giving a limit to the opening
angle of the scattering cone. In analogy to the far field
diffraction angle of a Gaussian beam we find for a Gaussian source
distribution an opening angle $\nu_{tr}$ of:
\begin{eqnarray}
  \nu_{tr}\approx\tan \nu_{tr}=\frac{\lambda}{\pi w_{a}}
\end{eqnarray}
Narrow samples scatter thus more efficiently than wide samples and
integrating the angular distribution of scattered intensity
predicts a $1/w_{a}^2$ dependence on the transverse width of the
sample.

Next we consider a pencil-shaped atomic sample. For this sample
$L_{0}\gg w_{a}$. By dividing the atomic sample again into two
parts (Fig.\ref{fig:diff_cones}) we can estimate the angle at
which the longitudinal extent of the cloud causes destructive
interference. Introducing the path length difference $\delta$,
using a small angle approximation and taking into account the
gaussian apodization  we can estimate the opening angle
$\theta_{L}$ as follows:
\begin{eqnarray}
  \delta =L_{0}(1-\cos\theta_{L})\approx L_{0}\frac{\theta^{2}_{L}}{2}=\frac{\lambda}{2 \pi}\nonumber\\
  \theta_{L}=\left(\frac{\lambda}{\pi L_{0}}\right)^{1/2}
\end{eqnarray}

Equating the two expressions for the opening angle we can define a
characteristic length $z_{ra}$, the atomic Rayleigh range, to
compare the influence of the transverse and the longitudinal
extent:
\begin{equation}
  z_{ra} = \frac{\pi w_{a}^{2}}{\lambda}
\end{equation}
For atomic samples of length $L_{0}$ comparable or longer than
$z_{ra}$, scattered waves from different sections along the
propagation direction will be mismatched in phase and the total
scattering cross section will be significantly reduced with
respect to a short sample with the same number of atoms.
\begin{figure*}[!t]
  \centering
  \hfill
  \includegraphics[width=0.4\textwidth]{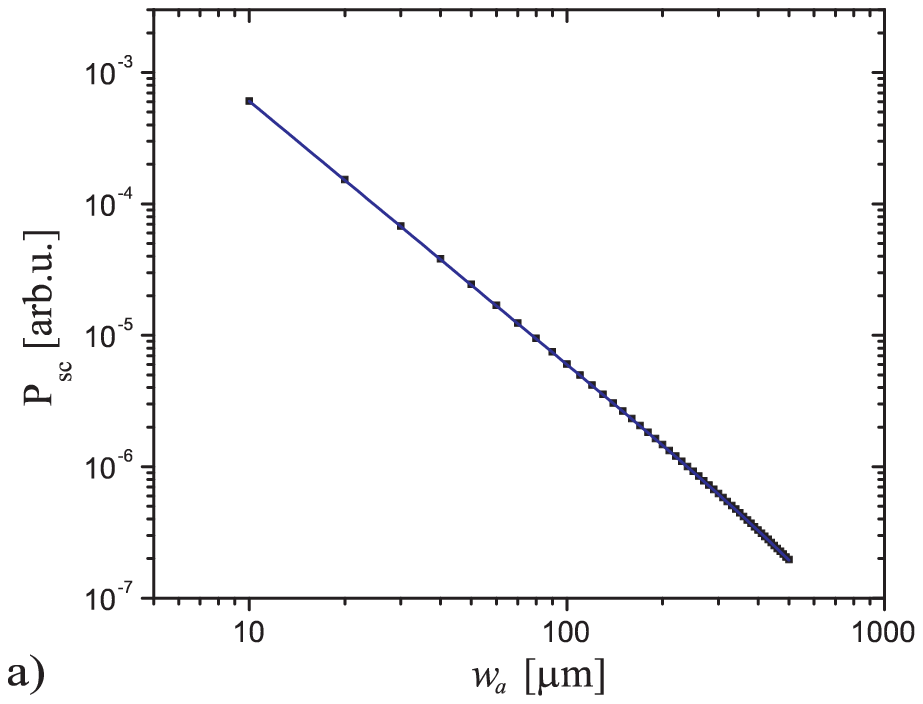}%
  \hfill
  \includegraphics[width=0.38\textwidth]{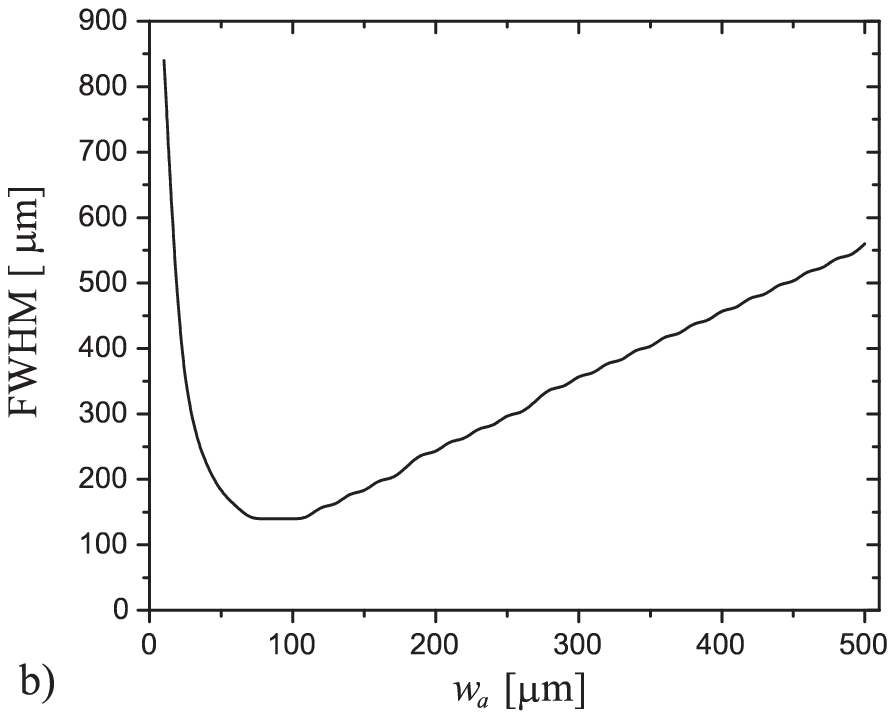}%
  \hspace*{\fill}
  \caption{(color online) a) Power of the scattered wave (symbols) vs.
   the characteristic transverse radius of atomic sample of
   length $L_{0}=1\mu m$ for constant number of atoms and a wide probe beam
   $w_{0} = 1000\mu m$ together with the analytic prediction from eq.(\ref{surprise})(solid line). b) FWHM of
   the intensity distribution in the observation plane for the same
   parameters.}
  \label{fig:power vs size}
  \end{figure*}

We can construct an approximate expression for the total scattered
power combining the above arguments in order to cast the influence
of the experimental parameters sample width, sample length and
beam diameter into a compact formula. Neglecting for a moment the
change of transverse spread over the length of the sample, the
scattered intensity on the optical axis far away from the sample
where all atoms are phase matched is approximately:
\begin{equation}
I_{sc}(0,0,z') \simeq \frac{3 \sigma N_{at}^{2}}{8 \pi z'^{2}}
\frac{2 P_{probe}}{\pi} \frac{w_{0}^{2}}{(w_{a}^{2} +
w_{0}^{2})^{2}}
\end{equation}
as can be verified easily by integration of eq.(\ref{E-scat}) in
the appropriate limit. To find the total scattered power we
replace the integration over the solid angle by a multiplication
with $\pi/2 \cdot \theta_{eff}^{2}$, where the effective opening
angle $\theta_{eff}$ is chosen with the help of the Fraunhofer
diffraction considerations from above. The extra factor of $1/2$
takes care of the very close to Gaussian profile of the scattered
wave. We expect this to be an excellent approximation whenever the
scattering cone is narrow. In order to model the tradeoff between
transversal and longitudinal limitation we have to design a
function which takes the value of the smaller of the two angles
whenever they are grossly different. We take:
\begin{equation} \label{theta-eff}
\theta_{eff} = (\frac{\theta_{T}^{2}
\theta_{L}^{2}}{(\theta_{T}^{4} + \theta_{L}^{4})^{1/2}})^{1/2}
\nonumber
\end{equation} with
\begin{equation}
  \theta_{T}^{2} = \frac{\lambda^{2}(w_{a}^{2} +
w_{0}^{2})}{\pi^{2} w_{0}^{2}w_{a}^{2}}
\end{equation}
Here the transverse limit angle takes into account diffraction
both due to the sample width as well as due to the probe beam
width. There is a great deal of freedom in the choice of
$\theta_{eff}$ and different definitions will lead to different
functional dependencies of the scattering efficiency on the length
of the sample. Our specific choice for $\theta_{eff}$ is motivated
by the crossover we observe in our numerical calculations for wide
probe beams presented below. Inserting the above formula we arrive
after some straightforward algebra at a compact expression for the
scattered power as:
\begin{equation} \label{surprise}
P_{sc} \simeq P_{probe} N_{at}^{2} \frac{ 3\sigma \lambda^{2}}{4
\pi^3 w_{0}^{2}w_{a}^{2}} \frac{1}{1 + w_{a}^{2}/w_{0}^{2}}
\frac{1}{\sqrt{1 + (L_{0}/\tilde{z}_{ra})^2}}
\end{equation}
Here we introduced $\tilde{z}_{ra}$, the modified atomic Rayleigh
range by using the definition of $\theta_{T}$ from
eq.\ref{theta-eff} and the relation $\tilde{z}_{ra} =
\lambda/(\pi\theta_{T}^{2})$.
\subsection{Numerical solutions of the diffraction integral}

Armed with the intuitive arguments and  eq.(\ref{surprise}) for
the scattering efficiency we can now proceed to present some of
our numerical results. We choose parameters for the D$_{2}$ line
of atomic Cs in our calculations ($\lambda = 852 nm$) and keep
probe power, detuning and number of atoms fixed for all results
presented in this section. Fig.\ref{fig:power vs size}a shows the
total scattered power in the observation plane for short samples
of varying transverse size.

The samples are probed by a wide
($w_{0}=1000\mu m$) probe beam. The scattering efficiency drops
dramatically with increasing sample size as expected. A comparison
with the $1/w_{a}^2 \cdot 1/(w_{a}^{2} + w_{0}^{2})$ dependence
from our analytical estimate shows perfect agreement. The full
width at half maximum (FWHM) of the intensity distribution in the
observation plane (Fig.\ref{fig:power vs size}b) reflects the
interplay of source size and diffraction in the propagation of the
scattered wave. In fact, our observation plane is not located in
the true far field for all source sizes and the observed
dependence is equivalent to the behavior of the spread of varying
size Gaussian beams at a fixed finite distance from their minimum
waist position.

\begin{figure*}[!t]
  \centering
  \hfill
  \includegraphics[width=0.46\textwidth]{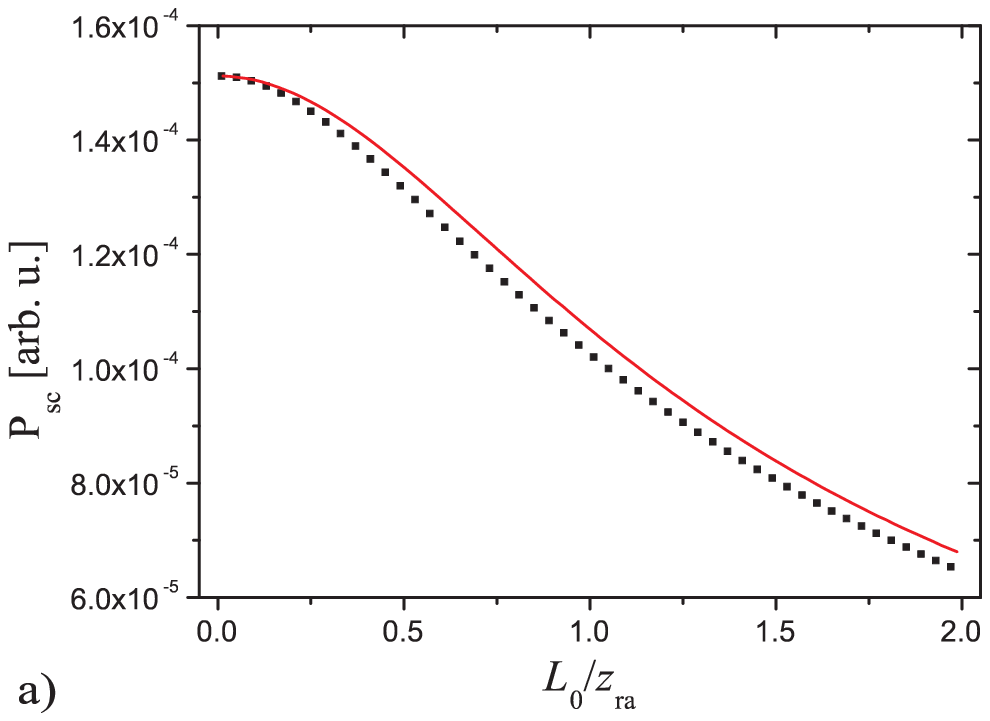}%
  \hfill
  \includegraphics[width=0.43\textwidth]{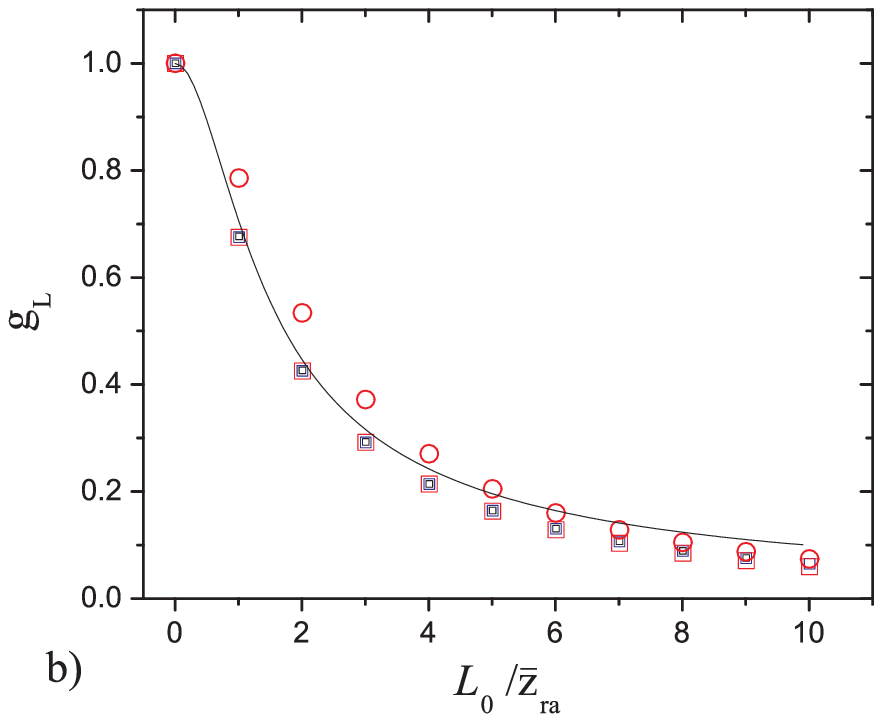}%
  \hspace*{\fill}
  \caption{(color online) a) Scattered power vs.
   the characteristic length of the atomic sample with
   atomic waist radius $w_{a} = 20\mu m$ probed by a beam $w_{0} = 1000\mu m$.
   Numerical data (symbols) and analytic prediction from eq.(\ref{surprise}) (solid line) are shown
   together. b) Same as in a) for sample width $w_{a} = 3, 5, 10, 20\mu
   m$ (squares) and for a narrow probe beam $w_{0} = w_{a} = 20\mu
   m$ (circles) with the length scaled to $\tilde{z}_{ra}$.
    }
   \label{fig:graph_lengthscan}
\end{figure*}

We investigate now how the scattered power changes with the length
of the sample. In Fig.\ref{fig:graph_lengthscan}a we show the
result for a narrow sample probed by a wide beam together with the
analytical prediction as a function of sample length in units of
$z_{ra}$. The rather good agreement with the simple function was
our motivation to design the expression for $\theta_{eff}$
accordingly. In Fig.\ref{fig:graph_lengthscan}b we repeat the
calculation over a larger range of scaled length for various
transverse sizes of the sample for wide (open squares in
Fig.\ref{fig:graph_lengthscan}b) and narrow (open circles in
Fig.\ref{fig:graph_lengthscan}b) probe beam. The length of atomic
sample is scaled here in units of $\tilde{z}_{ra}$ and the
scattered power is normalized to its value at infinitesimally
short sample length.

The estimate with our simple analytical formula is reasonable also
over this larger range showing quantitative agreement at the level
of 20\% for scaled sample length up to $(L_{0}/\tilde{z}_{ra})=8$.
The fact, that scaled data for a large probe beam diameter agree
among each other much better than with data from a small beam
diameter is understandable from the way eq.(\ref{surprise}) was
derived, i.e. neglecting explicitly the change of the probe beam
geometry over the length of the sample. To understand the effects
of changing probe geometry we consider the sample cut into thin
slices. We can identify each slice as a source for a Gaussian beam
wavelet, which initially inherits the phase profile of the probe
beam and develops wavefront curvature upon propagation. Scattered
wavefronts from the back end and the front end of the sample will
thus have different curvature limiting the overlap to small angles
around the forward direction. Our analytic formula works well for
a plane probe wavefront, but fails to take into account the
positive effect of a focused probe beam, which imprints wavefront
curvature of the right sign to enhance wavefront overlap. For this
reason the scaled scattering efficiency is slightly higher for
narrow probe beams and intermediate sample length than the model
predicts. For very elongated samples we see a systematic deviation
of the analytical model from the numerical result and going to
extremely elongated geometries we observe a different power-law of
the decay than suggested by the simple analytic model (see
Appendix A).  We refrain here from further tuning of the analytic
model, since this limiting case is of little interest for coupling
\textit{all} atoms of a realistic sample efficiently to the light
field and accurate quantitative data can be obtained easily
numerically whenever needed, anyway. We define the geometric
factor $g_{L}$ as the function describing universally the length
dependence:
\begin{equation}\label{gL}
 g_{L} = P_{sc}(L_{0})/P_{sc}(0) \simeq
 \frac{1}{\sqrt{1+(L_{0}/\tilde{z}_{ra})^{2}}}
\end{equation}

\begin{figure*}[!t]
  \centering
  \hfill
  \includegraphics[width=0.4\textwidth]{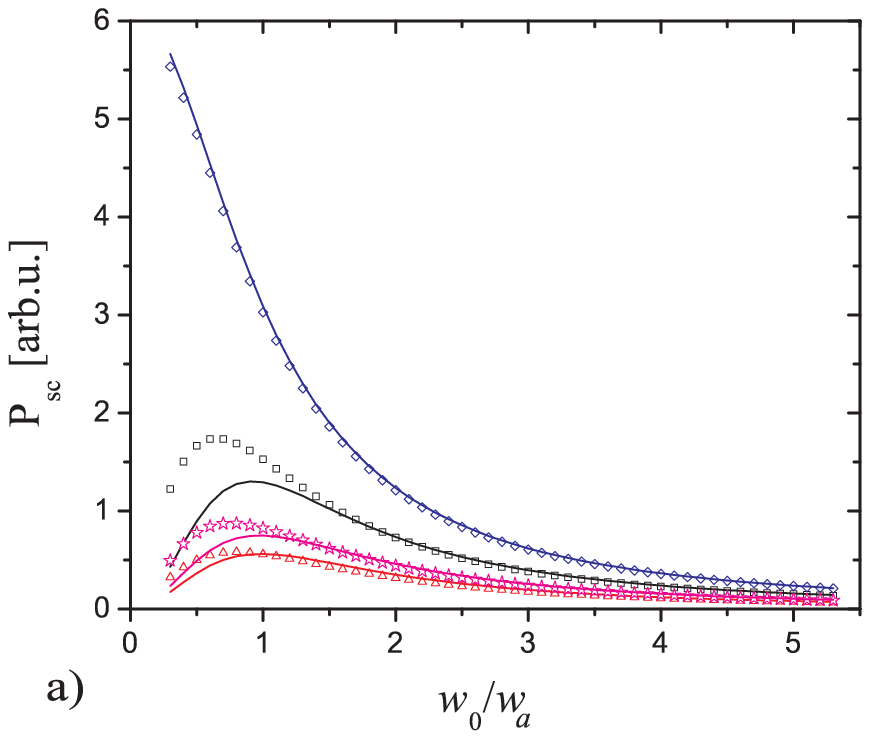}%
  \hfill
  \includegraphics[width=0.42\textwidth]{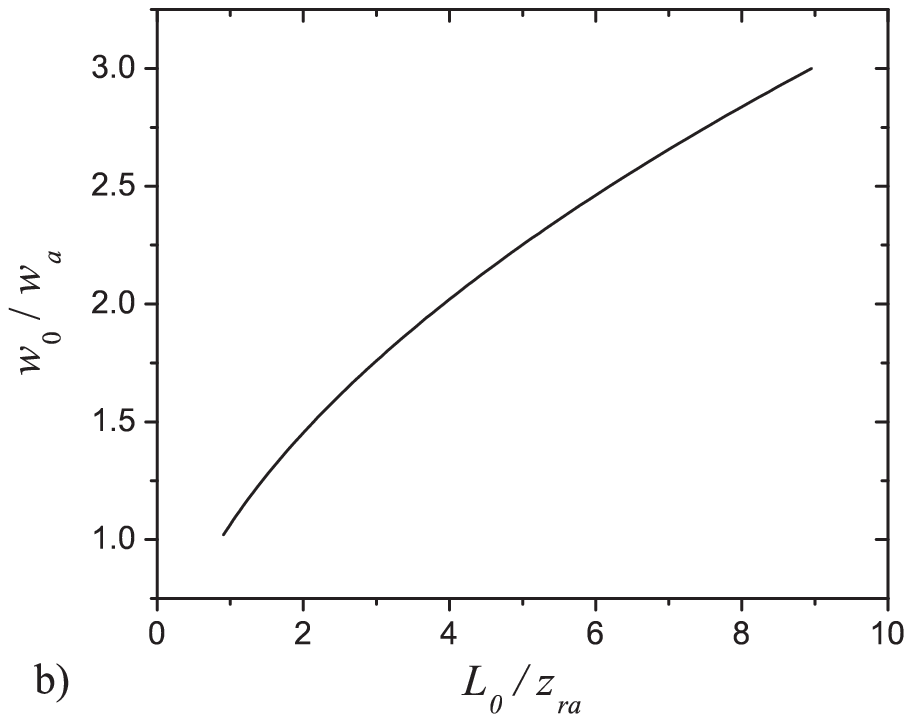}%
  \hspace*{\fill}
  \caption{(color online) a) Scattered power vs relative size of sample and probe
   beam for samples ($w_{a} = 10 \mu m$) of length $L_{0}=1,400,738,1000 \mu m$ (diamonds, squares, stars, triangles)
   together with the analytic prediction from eq.(\ref{surprise})(solid lines) in scaled units, b) Predicted probe beam size to match
   input and scattered wave as a function of sample length in scaled units.}
  \label{fig:probe beam size}
\end{figure*}

The third important parameter which can be varied in a real
experiment is the probe beam size. A probe beam size very much
larger than the sample size will not be optimum. While the sample
is illuminated homogeneously the field strength experienced by the
atoms is rather low. The dependence with decreasing probe diameter
predicted by eq.(\ref{surprise}) is the result of a subtle
interplay of increased single atom scattering at higher
intensities, increased coherent scattering efficiency for the
subset of atoms which is inside the volume covered by the probe
beam and decrease of the number of scatterers contributing
effectively to the scattered field. In Fig.\ref{fig:probe beam
size}a we show the scattered power as a function of probe beam
size for various sample length. Again we find that the transverse
probe beam size dependence is described very well by the analytic
formula for the case of a short sample. For short samples it is
advantageous to use a probe beam size as small as possible. For
longer samples both the numerical calculation as well as the
analytical estimate predict a finite probe beam diameter for
optimum scattering efficiency. For reasons already discussed in
the context of Fig.\ref{fig:graph_lengthscan}b the analytical
estimate starts to deviate from the numerical result for probe
diameters comparable or less than the sample diameter, but works
very well already for ratios as small as $2$. We separate the
trivial dependence on probe intensity and degree of transverse
localization ($\propto w_{0}^{-2} \cdot w_{a}^{-2}$) from the
observed behavior and define a geometric factor $g_{T}$ describing
the influence of the ratio of beam size to sample size
$w_{0}/w_{a}$ on the sample scattering efficiency as:
\begin{equation}\label{eq:transverse_geom_factor}
    g_{T}= \frac{1}{1 + w_{a}^{2}/w_{0}^{2}}
\end{equation}

For the polarization interferometric setup in
Fig.\ref{fig:interferometer}b it is not possible to adapt the
reference wavefront to the scattered wavefront separately, so
instead of maximizing the scattering efficiency only, one needs to
choose the input beam size, such that the scattered mode has also
good overlap with the input mode. For short samples it is easy to
see, that this can be achieved only with a probe size much smaller
than the sample width, since in this limit the probe traverses a
homogeneous region of the sample. For longer samples the
scattering cone narrows and one can achieve good overlap also for
a probe size comparable to the sample width. Equating the
far-field diffraction angle of the input beam with the effective
diffraction angle for light scattered off the sample, we can
derive an expression for the sample length which approximately
matches input and output modes:
\begin{equation}
\left(\frac{L_{0}}{z_{ra}}\right)^{2} =
\left(\frac{w_{0}}{w_{a}}\right)^{4}
\frac{\left(1+\left(\frac{w_{0}}{w_{a}}\right)^{2}\right)^{2}
-1}{\left(1+\left(\frac{w_{0}}{w_{a}}\right)^{2}\right)^{2}}
\end{equation}
The interesting region for the ratio $w_{0}/w_{a}$ is values
bigger than 1, i.e. probe sizes comparable or bigger than the
sample size. The predictions of the above equation are shown
graphically in Fig.\ref{fig:probe beam size}b. The values obtained
from the analytical formula provide good starting values for a
numerical optimization of this mode-matching problem.

 With our numerical calculations we explored the range of validity of a
simple analytical estimate for the scattering efficiency from
samples of different size and found quantitative agreement at the
20\% level over a large range of parameters.

\begin{figure*}[!t]
  \centering
  \hfill
  \includegraphics[width=0.4\textwidth]{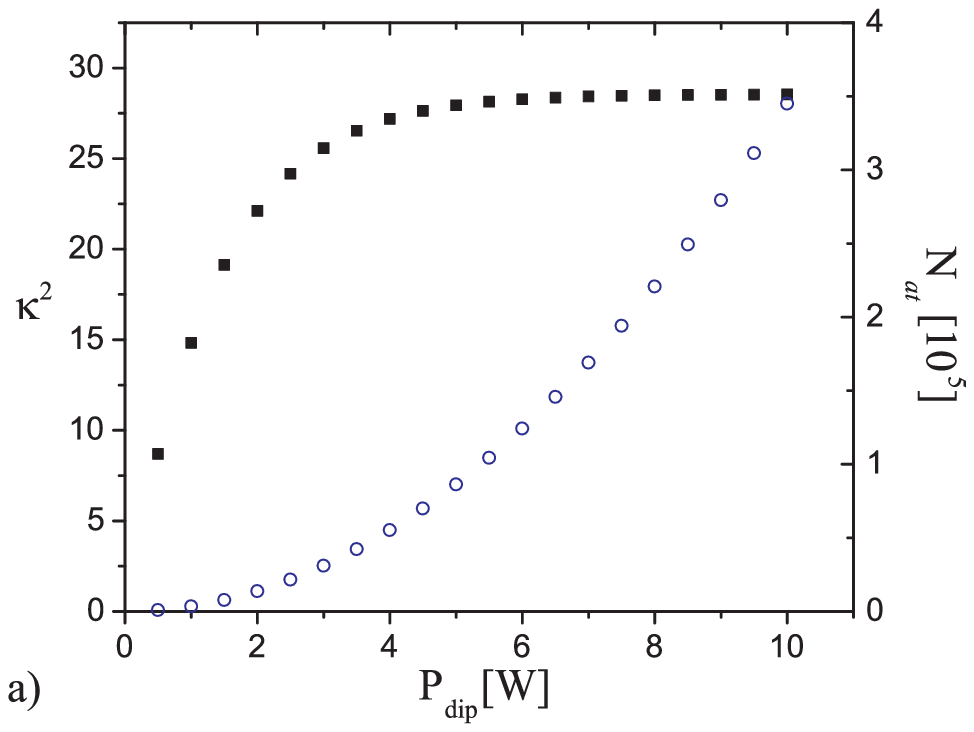}%
  \hfill
  \includegraphics[width=0.4\textwidth]{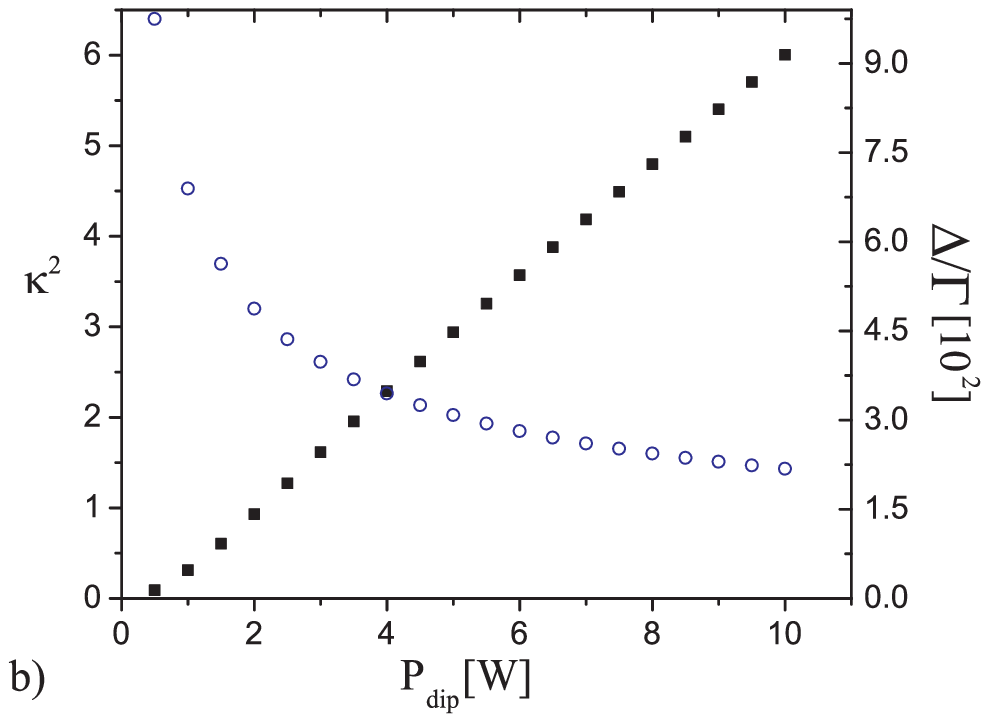}%
  \hspace*{\fill}
  \caption{(color online) a) achievable coupling strength (filled symbols, left axis) and number of trapped
  atoms(open symbols, right axis) as a function of invested dipole
  trap power; b) achievable coupling strength (filled symbols,
  left axis) and probe detuning (open symbols, right axis) needed to
satisfy $\eta = 0.1$
  (see text for details).}\label{fig:SNRvsPdip}
\end{figure*}

\section{Application to a dipole trapped sample}

The formula and numerical calculation suggest that a global
optimum for the scattering efficiency exists for any sample, which
simply consists in placing all scatterers in one point in space.
This optimum is, alas, unphysical, because dipole-dipole
interaction in that case dominates the scattering physics and,
more pragmatically, a trapped sample is subject to density
limitations because of collision induced heating and losses. In
the following we study the case of atoms trapped in a single
gaussian beam dipole trap. In thermal equilibrium the shape of the
atomic sample in a single beam dipole trap is determined by the
focal parameter of the beam \cite{dipole, Footnote10}. We take
parameters for Cs atoms trapped by laser radiation at
$\lambda_{dip} = 1030 nm$ at a constant trap depth of $U_{0} =
k_{B}\cdot 1 mK$ and fixed sample temperature of $T = 100 \mu K$.
Specifying the dipole trap laser power determines then the focal
parameter needed to achieve the trap depth, and thus also the
thermal radius $w_{a}$ and length $L_{0}$ of the sample. Limiting
the peak density to $n_{peak} = 10^{12} cm^{-3}$ specifies then
the number of atoms $N_{at}$. From Fig.\ref{fig:probe beam size}a
we infer that for long samples a probe beam size equal to the
sample size will be close to optimum and choose this for the
calculation. Restoring all prefactors and choosing a number of
incident probe photons $n_{ph} = 10^8$ \cite{Footnote11} at a
detuning $\Delta/\Gamma = 100$ we can then numerically determine
the achievable SNR according to eq.(\ref{eq:SNR+at_fluct}) at
unity quantum efficiency as a function of the power of the dipole
trap laser. In Fig.\ref{fig:SNRvsPdip}a we show the achievable SNR
in this configuration, together with the number of trapped atoms.
We observe that for bigger samples the SNR approaches a constant
value. With the constraints we placed on temperature and density,
the benefit of having more atoms is reduced by the increasingly
unfavorable elongated geometry. It turns out that while the aspect
ratio of the sample increases with increasing size of the dipole
trap, the scaled length, i.e. the Fresnel number, remains constant
at $L_{0}/\tilde{z}_{ra} = 14$ \cite{Footnote16}. With this
observation, the higher number of atoms is outweighed exactly by
the increasing transverse dimensions of sample and probe beam.

A realistic model for coherent light-atom coupling efficiency will
have to take into account also the losses due to spontaneous
emission. In fact, a measurement on atoms with a spontaneous
emission probability approaching $1$ can hardly be considered
non-destructive for the collective variable. The result from the
1-D quantum model in eq.(\ref{SignaltoNoise1-D}) predicts actually
that the achievable SNR and the level of destructiveness are
coupled, i.e. the achievable SNR is directly proportional to the
integrated single atom spontaneous emission rate $\eta$ (see also
\cite{Jessen, Lye}). Since the transverse size of the trapped
sample changes with the invested dipole trap power, $\eta$ is not
the same for the data points in Fig.\ref{fig:SNRvsPdip}a. In order
to make a fair comparison we thus calculate for each of the data
points the number of spontaneously emitted photons per atom by
evaluating the average intensity experienced by the atoms and
integrating over the pulse time (see Appendix B). In
Fig.\ref{fig:SNRvsPdip}b we show the same data rescaled to a probe
laser detuning, such that $\eta = 0.1$ for every point, together
with the detuning needed to satisfy the constraint on $\eta$.

Including the condition of equal level of destructiveness restores
the advantage of bigger samples over small samples. We note that
the absolute numbers for the coupling strength at fixed $\eta$
cannot be changed by simultaneous variation of the detuning and
the number of incident photons, since the coupling strength and
$\eta$ depend in the same way on these two quantities. For the
elongated samples probed by a narrow beam the losses are
distributed quite unevenly over the sample due to the rather
inhomogeneous illumination.

\section{Relation to effective 1-D models}
\label{relation}
 At the start of our scattering calculation we
expressed the coupling strength or achievable SNR in terms of a
number of coherently scattered photons. This number, although
convenient to calculate, is not a directly measurable quantity,
since we cannot distinguish coherently scattered photons from the
incident photons in principle and only their interference is
observable. The results from the previous sections allow us to
express this artificial scattered power in terms of the incident
power and the interaction geometry. This also makes a direct
comparison to the expression for the coupling strength derived
from the 1-D quantum model possible. Introducing the transverse
beam area $A_{ph}$ as $\pi w_{0}^2$ and equivalently the sample
area $A_{at}$ as $\pi w_{a}^2$ we rewrite eq.(\ref{surprise}) in
the limit of large detuning as \cite{Footnote12}:
\begin{eqnarray}\nonumber
   P_{sc}= \frac {3}{2} N_{at}^{2}\frac{\sigma_{0}}{A_{ph}}\frac{\sigma_{0}}{A_{at}}\left(\frac{\Gamma}{2\Delta}\right)^{2}g_{T}g_{L}P_{inc}
\end{eqnarray}

 Using eq.(\ref{eq:SNR+at_fluct}) we obtain the SNR assuming unit quantum efficiency detection as:
\begin{eqnarray}\label{eq:SNR_geom_factors}
   \left(\frac{S}{N}\right)^{2}=  g_{T} g_{L}\frac{3\sigma_{0}^{2}}{A_{at}A_{ph}}N_{at}n_{ph}\left(\frac{\Gamma}{2\Delta}\right)^{2}
\end{eqnarray}

Comparing this to the expression obtained from an effective 1-D
model in section \ref{collective}:
\begin{eqnarray}\label{eq:signal/noise_1D}
 \kappa^{2}=
 \frac{\sigma_{0}^{2}}{A^{2}}N_{at}n_{ph}\left(\frac{\Gamma}{2\Delta}\right)^{2},
\end{eqnarray}
we see how diffraction effects modify the coupling strength with
respect to the predictions from the 1-D model. The two expressions
are similar for the case of a sample with effective Fresnel number
of $1$ and equal probe and sample diameter \cite{Raymer}. For our
choice of Gaussian sample and probe the modification of coupling
strength can be quantitatively accounted for by the geometrical
factors. We believe that the asymptotic scaling of the geometrical
factors will be independent of the specific choice of the function
describing the shape of the sample. We can look at the case of a
very elongated sample ($L_{0}\gg \tilde {z}_{ra}$) probed by a
narrow beam of the same size as the sample by expanding the
geometric factors accordingly and find in this limit for the
coupling strength:
\begin{eqnarray}\label{eq:SNR_long}
   \kappa^{2}=\frac{\pi ^{3/2}}{8}\left(\frac{\lambda}{2\pi}\right)^{3}n_{peak}n_{ph}\left(\frac{\Gamma}{2\Delta}\right)^{2}.
\end{eqnarray}
The achievable coupling strength becomes independent of the sample
size in this limit and is linear in the peak atomic density
$n_{peak}$, instead of being proportional to column density as in
the 1-D model \cite{Footnote13}. The numerically observed scaling
with the length of the sample (see the appendix) reduces the
coupling even more for extremely elongated samples.

\section{Conclusion}
In this paper we have outlined an efficient method to include
diffraction effects in the coupling of light to collective
variables of atomic samples and applied it to an experimentally
relevant case of atomic ensembles stored in single beam dipole
traps. The use of gaussian light fields is well adapted to real
experimental geometries and allows for a largely analytical
treatment. Tayloring the sample and beam geometry, such that probe
mode and scattered mode coincide is possible and will be useful
for polarization interferometry or multi-pass experiments.

Several approximations have been made, mainly to keep the model as
transparent as possible, and some of them can be lifted in future
extensions of the model. The leading order effect of multiple
scattering and particle statistics on the refractive index, which
we are effectively calculating in a single scattering
approximation, can be accounted for by a correction term depending
on the local density of scatterers which will allow to calculate
the geometry of the scattered mode also for higher densities \cite{Morice}.\\

Our model is classical in nature, but the point scatterer model
can be used also to analyze quantum noise contributions and their
dependence on geometry. Giving up the continuous density
distribution one can determine numerically the scattering
efficiency from randomly distributed samples and this way
statistically analyze the noise introduced onto the scattered
wavefront by density fluctuations on different length scales. This
models spontaneous emission noise as well as nontrivial effects
like the inherent mode-matching noise discussed in
\cite{LuMingDuan}. There are already studies using a wave function
Monte-Carlo technique to address the effects of spontaneous
emission in a quantum description \cite{Bouchoule} adapted for our
case of trapped inhomogeneous samples, but the extremely fast
increase of the dimensionality of Hilbert space limits the
treatment to very small numbers of atoms only. A classical point
scatterer calculation can be used conveniently also to model
experimental imperfections, e.g. alignment errors, where the
analytical integration over the transverse distribution would
become much more involved.

 The assumption of infinitely heavy
scatterers, i.e. the neglect of photon recoil, needs closer
attention, when the fluctuations of collective variables are
studied. In fact, when working with collective atomic variables
one usually assumes that internal and external degrees of freedom
of the atomic sample are decoupled. Already without multiple
scattering the change of momentum due to scattering introduces
correlations between internal and external variables
\cite{Burnett}. Also, a focused probe beam with inhomogeneous
intensity distribution across the sample exerts a dipole force on
the sample leading to contraction or expansion depending on the
sub-level populations for the setup in
Fig.\ref{fig:interferometer}a. This leads to an effective decay
mechanism for the macroscopic coherence between the sub-levels.
Similar effects occur naturally also at the level of quantum
fluctuations. Ultimately, a proper quantum model will have to take
into account the scattering induced dynamics of the density
correlation function of the sample, which determines the structure
factor for light scattering \cite{Footnote14, Politzer}. Prominent
examples for the key importance of the photon recoil for
collective scattering are the observation of super-radiant
Rayleigh and Raman scattering in Bose-Einstein condensates
\cite{Ketterle1, Ketterle2}, cavity cooling \cite{Horak, Vuletic}
and collective motion in high-finesse cavities \cite{Hemmerich}.

\begin{acknowledgments}
J.H.M. thanks K.M{\o}lmer and A.S{\o}rensen for stimulating
discussions. We acknowledge support for this work by the
EU-network CAUAC and the Dansk Grundforskningsfond.
\end{acknowledgments}

\section{Appendix}
\subsection{}
The analytical estimate for the total scattered power given in
eq.(\ref{surprise}) is seen to fail for very elongated samples.
This can be traced back to the assumption of homogeneous
illumination in the calculation of the on-axis intensity of the
scattered field. A simple way to arrive at an improved estimate is
to introduce an axial average of the incident intensity in order
to take into account the diffractive spreading of the incident
beam over the sample length. Since the scattering efficiency
depends quadratically on atom number, the average is performed
over the squared density distribution and to simplify the math the
gaussian atomic density distribution is replaced by a rectangular
distribution of same peak height and area:
\begin{eqnarray}
I_{eff} = \frac{I_{0}}{\sqrt{\pi/2}L_{0}}
\int_{0}^{\sqrt{\pi/2}L_{0}} \frac{1}{1+(z/z_{r})^2} dz & &
\nonumber \\
 = I_{0}\frac{z_{r}}{\sqrt{\pi/2}L_{0}}
\arctan(\frac{\sqrt{\pi/2}L_{0}}{z_{r}}).
\end{eqnarray}
The inhomogeneous axial illumination changes also the effective
length of the sample entering the estimate for the opening angle
of the diffraction cone. Incorporating this effect we find
empirically an improved expression for the longitudinal
geometrical factor (see eq.(\ref{gL}))
\begin{eqnarray}
g'_{L}& = & \frac {z_{r}}{\sqrt{\pi/2} L_{0}} \arctan
\left\{\frac{\sqrt{\pi/2} L_{0}}{z_{r}}
\left(\frac{1+\left(\frac{L_{0}\tilde{z}_{ra}}{z_{r}^2}\right)^{2}}
{1+\left(\frac{L_{0}}{\tilde{z}_{ra}}\right)^2}\right)^{1/2}\right\}
\nonumber \\
& & \times
\frac{1}{\sqrt{1+\left(\frac{L_{0}\tilde{z}_{ra}}{z_{r}^2}\right)^2}}
\end{eqnarray}
which fits our numerical data at the level of 20\% for Fresnel
numbers of the atomic sample up to 80.
\subsection{}
Within the framework of the point scatterer model the distinction
between spontaneous and induced emission is blurred and with the
approximation of a microscopically continuous density distribution
for the point scatterers spontaneous emission is completely lost.
Introducing implicitly microscopic density fluctuations and
assuming that single atom spontaneous emission happens
independently of the presence of the neighboring scatterers the
spontaneously emitted power can be calculated as:
\begin{equation}
\langle P_{spon} \rangle = N_{at} \sigma \int_{\mathbb{R}^3}
I_{inc}(\vec{r}) n(\vec{r}) dV
\end{equation}
Setting for simplicity the wavelength of the dipole trap laser,
which determines the change in transverse dimensions of the atomic
sample, equal to the wavelength of the incident radiation, the
single atom spontaneous emission rate can be written in the form:
\begin{eqnarray}
\frac{\eta}{\tau} = \frac{\langle P_{spon} \rangle} {N_{at}} = &
\sigma \frac{2 P_{probe}}{\pi w_{0}^2}
\frac{1}{1+2(w_{a}/w_{0})^2} \pi^{-1/2}& \nonumber \\
& \times\int_{-\infty}^{\infty} \frac{1}{1+(z/a)^2} \exp(-z^2) dz
\end{eqnarray}
with
\begin{equation}\nonumber
a = \frac{z_{r}}{L_{0}}\left(\frac{1 + 2(w_{a}/w_{0})^2}{3}
\right)^{1/2}
\end{equation}
and
\begin{equation}\nonumber
\int_{-\infty}^{\infty} \frac{1}{1+(z/a)^2} \exp(-z^2) dz = a \pi
\exp(a^2) Erfc(a)
\end{equation}
where $Erfc(a)$ denotes the complementary error function. The
separation of the expression for the coupling strength $\kappa^2$
into integrated spontaneous emission $\eta$ and an effective
optical depth $\alpha$ as in the 1-D description can be done in
principle, but does not lead to simple analytical expressions. In
fact, such a separation is also not very meaningful when done
globally, since the integrated spontaneous emission rate can have
substantial \textit{local} variations due to the inhomogeneous
illumination. In addition, the contribution of a central volume
element to the total scattered field and thus to the scattering
efficiency is much bigger than for a volume element on the rim of
the density distribution. This means that a spontaneous emission
or optical pumping event in the center of the sample leads to a
more pronounced change in the scattering efficiency. In a quantum
description the spatial inhomogeneity of both light and atom
variables naturally suggests importance sampling and leads to a
concept of collective variables which are no longer fully
symmetric with respect to exchange of single particle labels
\cite{Kuzmich04}.

\bibliographystyle{apsrev}
\bibliography{diff_ref}

\begin{thebibliography}{50}
\expandafter\ifx\csname natexlab\endcsname\relax\def\natexlab#1{#1}\fi
\expandafter\ifx\csname bibnamefont\endcsname\relax
  \def\bibnamefont#1{#1}\fi
\expandafter\ifx\csname bibfnamefont\endcsname\relax
  \def\bibfnamefont#1{#1}\fi
\expandafter\ifx\csname citenamefont\endcsname\relax
  \def\citenamefont#1{#1}\fi
\expandafter\ifx\csname url\endcsname\relax
  \def\url#1{\texttt{#1}}\fi
\expandafter\ifx\csname urlprefix\endcsname\relax\def\urlprefix{URL }\fi
\providecommand{\bibinfo}[2]{#2}
\providecommand{\eprint}[2][]{\url{#2}}

\bibitem[{\citenamefont{Kuzmich et~al.}(1997)\citenamefont{Kuzmich, M{\o}lmer,
  and Polzik}}]{KuMoPo}
\bibinfo{author}{\bibfnamefont{A.}~\bibnamefont{Kuzmich}},
  \bibinfo{author}{\bibfnamefont{K.}~\bibnamefont{M{\o}lmer}},
  \bibnamefont{and} \bibinfo{author}{\bibfnamefont{E.~S.}
  \bibnamefont{Polzik}}, \bibinfo{journal}{Phys. Rev. Lett.}
  \textbf{\bibinfo{volume}{79}}, \bibinfo{pages}{4782} (\bibinfo{year}{1997}).

\bibitem[{\citenamefont{Kuzmich et~al.}(1998)\citenamefont{Kuzmich, Bigelow,
  and Mandel}}]{KuzmichEuro}
\bibinfo{author}{\bibfnamefont{A.}~\bibnamefont{Kuzmich}},
  \bibinfo{author}{\bibfnamefont{N.~P.} \bibnamefont{Bigelow}},
  \bibnamefont{and} \bibinfo{author}{\bibfnamefont{L.}~\bibnamefont{Mandel}},
  \bibinfo{journal}{Europhys. Lett.} \textbf{\bibinfo{volume}{42}},
  \bibinfo{pages}{481} (\bibinfo{year}{1998}).

\bibitem[{\citenamefont{Duan et~al.}(2000)\citenamefont{Duan, Cirac, Zoller,
  and Polzik}}]{duan}
\bibinfo{author}{\bibfnamefont{L.~M.} \bibnamefont{Duan}},
  \bibinfo{author}{\bibfnamefont{J.~I.} \bibnamefont{Cirac}},
  \bibinfo{author}{\bibfnamefont{P.}~\bibnamefont{Zoller}}, \bibnamefont{and}
  \bibinfo{author}{\bibfnamefont{E.~S.} \bibnamefont{Polzik}},
  \bibinfo{journal}{Phys. Rev. Lett.} \textbf{\bibinfo{volume}{85}},
  \bibinfo{pages}{5643} (\bibinfo{year}{2000}).

\bibitem[{\citenamefont{Kuzmich and Polzik}(2000)}]{Kuzmich2000}
\bibinfo{author}{\bibfnamefont{A.}~\bibnamefont{Kuzmich}} \bibnamefont{and}
  \bibinfo{author}{\bibfnamefont{E.~S.} \bibnamefont{Polzik}},
  \bibinfo{journal}{Phys. Rev. Lett.} \textbf{\bibinfo{volume}{85}},
  \bibinfo{pages}{5639} (\bibinfo{year}{2000}).

\bibitem[{\citenamefont{Kozhekin et~al.}(2000)\citenamefont{Kozhekin,
  M{\o}lmer, and Polzik}}]{Kozhekin}
\bibinfo{author}{\bibfnamefont{A.~E.} \bibnamefont{Kozhekin}},
  \bibinfo{author}{\bibfnamefont{K.}~\bibnamefont{M{\o}lmer}},
  \bibnamefont{and} \bibinfo{author}{\bibfnamefont{E.}~\bibnamefont{Polzik}},
  \bibinfo{journal}{Phys. Rev. A} \textbf{\bibinfo{volume}{62}},
  \bibinfo{pages}{033809} (\bibinfo{year}{2000}).

\bibitem[{\citenamefont{Thomsen et~al.}(2002)\citenamefont{Thomsen, Mancini,
  and Wiseman}}]{wiseman}
\bibinfo{author}{\bibfnamefont{L.~K.} \bibnamefont{Thomsen}},
  \bibinfo{author}{\bibfnamefont{S.}~\bibnamefont{Mancini}}, \bibnamefont{and}
  \bibinfo{author}{\bibfnamefont{H.~M.} \bibnamefont{Wiseman}},
  \bibinfo{journal}{Phys. Rev. A} \textbf{\bibinfo{volume}{65}},
  \bibinfo{pages}{061801} (\bibinfo{year}{2002}).

\bibitem[{\citenamefont{Kraus et~al.}(2003)\citenamefont{Kraus, Hammerer,
  Giedke, and Cirac}}]{Kraus}
\bibinfo{author}{\bibfnamefont{B.}~\bibnamefont{Kraus}},
  \bibinfo{author}{\bibfnamefont{K.}~\bibnamefont{Hammerer}},
  \bibinfo{author}{\bibfnamefont{G.}~\bibnamefont{Giedke}}, \bibnamefont{and}
  \bibinfo{author}{\bibfnamefont{J.~I.} \bibnamefont{Cirac}},
  \bibinfo{journal}{Phys. Rev. A} \textbf{\bibinfo{volume}{67}},
  \bibinfo{pages}{042314} (\bibinfo{year}{2003}).

\bibitem[{\citenamefont{Fiur{\'{a}}{\u{s}}ek}(2003)}]{Fiurasek}
\bibinfo{author}{\bibfnamefont{J.}~\bibnamefont{Fiur{\'{a}}{\u{s}}ek}},
  \bibinfo{journal}{Phys. Rev. A} \textbf{\bibinfo{volume}{68}},
  \bibinfo{pages}{022304} (\bibinfo{year}{2003}).

\bibitem[{\citenamefont{Hald et~al.}(1999)\citenamefont{Hald, S{\o}rensen,
  Schori, and Polzik}}]{Hald}
\bibinfo{author}{\bibfnamefont{J.}~\bibnamefont{Hald}},
  \bibinfo{author}{\bibfnamefont{J.~L.} \bibnamefont{S{\o}rensen}},
  \bibinfo{author}{\bibfnamefont{C.}~\bibnamefont{Schori}}, \bibnamefont{and}
  \bibinfo{author}{\bibfnamefont{E.~S.} \bibnamefont{Polzik}},
  \bibinfo{journal}{Phys. Rev. Lett.} \textbf{\bibinfo{volume}{83}},
  \bibinfo{pages}{1319} (\bibinfo{year}{1999}).

\bibitem[{\citenamefont{Kuzmich et~al.}(2000)\citenamefont{Kuzmich, Bigelow,
  and Mandel}}]{kuzmichprl}
\bibinfo{author}{\bibfnamefont{A.}~\bibnamefont{Kuzmich}},
  \bibinfo{author}{\bibfnamefont{N.~P.} \bibnamefont{Bigelow}},
  \bibnamefont{and} \bibinfo{author}{\bibfnamefont{P.}~\bibnamefont{Mandel}},
  \bibinfo{journal}{Phys. Rev. Lett.} \textbf{\bibinfo{volume}{85}},
  \bibinfo{pages}{1594} (\bibinfo{year}{2000}).

\bibitem[{\citenamefont{Julsgaard et~al.}(2001)\citenamefont{Julsgaard,
  Kozhekin, and Polzik}}]{briannature}
\bibinfo{author}{\bibfnamefont{B.}~\bibnamefont{Julsgaard}},
  \bibinfo{author}{\bibfnamefont{A.}~\bibnamefont{Kozhekin}}, \bibnamefont{and}
  \bibinfo{author}{\bibfnamefont{E.~S.} \bibnamefont{Polzik}},
  \bibinfo{journal}{Nature} \textbf{\bibinfo{volume}{413}},
  \bibinfo{pages}{400} (\bibinfo{year}{2001}).

\bibitem[{\citenamefont{Schori et~al.}(2002)\citenamefont{Schori, Julsgaard,
  S{\o}rensen, and Polzik}}]{Schori}
\bibinfo{author}{\bibfnamefont{C.}~\bibnamefont{Schori}},
  \bibinfo{author}{\bibfnamefont{B.}~\bibnamefont{Julsgaard}},
  \bibinfo{author}{\bibfnamefont{J.~L.} \bibnamefont{S{\o}rensen}},
  \bibnamefont{and} \bibinfo{author}{\bibfnamefont{E.~S.}
  \bibnamefont{Polzik}}, \bibinfo{journal}{Phys. Rev. Lett.}
  \textbf{\bibinfo{volume}{89}}, \bibinfo{pages}{057903}
  (\bibinfo{year}{2002}).

\bibitem[{\citenamefont{Duan et~al.}(2002)\citenamefont{Duan, Cirac, and
  Zoller}}]{LuMingDuan}
\bibinfo{author}{\bibfnamefont{L.~M.} \bibnamefont{Duan}},
  \bibinfo{author}{\bibfnamefont{J.~I.} \bibnamefont{Cirac}}, \bibnamefont{and}
  \bibinfo{author}{\bibfnamefont{P.}~\bibnamefont{Zoller}},
  \bibinfo{journal}{Phys. Rev. A} \textbf{\bibinfo{volume}{66}},
  \bibinfo{pages}{023818} (\bibinfo{year}{2002}).

\bibitem[{\citenamefont{Kuzmich and Kennedy}(2004)}]{Kuzmich04}
\bibinfo{author}{\bibfnamefont{A.}~\bibnamefont{Kuzmich}} \bibnamefont{and}
  \bibinfo{author}{\bibfnamefont{T.~A.~B.} \bibnamefont{Kennedy}},
  \bibinfo{journal}{Phys. Rev. Lett.} \textbf{\bibinfo{volume}{92}},
  \bibinfo{pages}{030407} (\bibinfo{year}{2004}).

\bibitem[{\citenamefont{Happer and Mathur}(1967)}]{Happer}
\bibinfo{author}{\bibfnamefont{W.}~\bibnamefont{Happer}} \bibnamefont{and}
  \bibinfo{author}{\bibfnamefont{B.~S.} \bibnamefont{Mathur}},
  \bibinfo{journal}{Phys. Rev. Lett.} \textbf{\bibinfo{volume}{18}},
  \bibinfo{pages}{577} (\bibinfo{year}{1967}).

\bibitem[{\citenamefont{Kuzmich et~al.}(1999)\citenamefont{Kuzmich, Mandel,
  Janis, Young, Ejnisman, and Bigelow}}]{Kuzmich1}
\bibinfo{author}{\bibfnamefont{A.}~\bibnamefont{Kuzmich}},
  \bibinfo{author}{\bibfnamefont{L.}~\bibnamefont{Mandel}},
  \bibinfo{author}{\bibfnamefont{J.}~\bibnamefont{Janis}},
  \bibinfo{author}{\bibfnamefont{Y.~E.} \bibnamefont{Young}},
  \bibinfo{author}{\bibfnamefont{R.}~\bibnamefont{Ejnisman}}, \bibnamefont{and}
  \bibinfo{author}{\bibfnamefont{N.~P.} \bibnamefont{Bigelow}},
  \bibinfo{journal}{Phys. Rev. A} \textbf{\bibinfo{volume}{60}},
  \bibinfo{pages}{2346} (\bibinfo{year}{1999}).

\bibitem[{Foo({\natexlab{a}})}]{Footnote1}
\bibinfo{note}{Here we drop terms which depend solely on the total number of
  atoms and photons.}

\bibitem[{Foo({\natexlab{b}})}]{Footnote2}
\bibinfo{note}{Strictly speaking the two setups are not equivalent, since
  vacuum is leaking into the interferometer in (a) before the interaction,
  while it couples into the polarizing beam splitter in (b) only after the
  interaction.}

\bibitem[{Foo({\natexlab{c}})}]{Footnote3}
\bibinfo{note}{Spatial correlations due to the indistinguishability of the
  particles are neglected here.}

\bibitem[{Foo({\natexlab{d}})}]{Footnote4}
\bibinfo{note}{This procedure is valid for Gaussian input states like coherent
  or squeezed states.}

\bibitem[{\citenamefont{Sinatra et~al.}(1998)\citenamefont{Sinatra, Roch,
  Vigneron, Grelu, Poizat, Wang, and Grangier}}]{Sinatra}
\bibinfo{author}{\bibfnamefont{A.}~\bibnamefont{Sinatra}},
  \bibinfo{author}{\bibfnamefont{J.~F.} \bibnamefont{Roch}},
  \bibinfo{author}{\bibfnamefont{K.}~\bibnamefont{Vigneron}},
  \bibinfo{author}{\bibfnamefont{P.}~\bibnamefont{Grelu}},
  \bibinfo{author}{\bibfnamefont{J.-P.} \bibnamefont{Poizat}},
  \bibinfo{author}{\bibfnamefont{K.}~\bibnamefont{Wang}}, \bibnamefont{and}
  \bibinfo{author}{\bibfnamefont{P.}~\bibnamefont{Grangier}},
  \bibinfo{journal}{Phys. Rev. A} \textbf{\bibinfo{volume}{57}},
  \bibinfo{pages}{2980} (\bibinfo{year}{1998}).

\bibitem[{\citenamefont{Wineland et~al.}(1994)\citenamefont{Wineland,
  Bollinger, Itano, and Heinzen}}]{Wineland}
\bibinfo{author}{\bibfnamefont{D.~J.} \bibnamefont{Wineland}},
  \bibinfo{author}{\bibfnamefont{J.~J.} \bibnamefont{Bollinger}},
  \bibinfo{author}{\bibfnamefont{W.~M.} \bibnamefont{Itano}}, \bibnamefont{and}
  \bibinfo{author}{\bibfnamefont{D.~J.} \bibnamefont{Heinzen}},
  \bibinfo{journal}{Phys. Rev. A} \textbf{\bibinfo{volume}{50}},
  \bibinfo{pages}{67} (\bibinfo{year}{1994}).

\bibitem[{\citenamefont{Hammerer et~al.}(2003)\citenamefont{Hammerer,
  M{\o}lmer, Polzik, and Cirac}}]{Hammerer}
\bibinfo{author}{\bibfnamefont{K.}~\bibnamefont{Hammerer}},
  \bibinfo{author}{\bibfnamefont{K.}~\bibnamefont{M{\o}lmer}},
  \bibinfo{author}{\bibfnamefont{E.~S.} \bibnamefont{Polzik}},
  \bibnamefont{and} \bibinfo{author}{\bibfnamefont{J.~I.} \bibnamefont{Cirac}}
  (\bibinfo{year}{2003}), \eprint{quant-ph/0312156}.

\bibitem[{\citenamefont{Kuzmich and Polzik}(2003)}]{Braunstein}
\bibinfo{author}{\bibfnamefont{A.}~\bibnamefont{Kuzmich}} \bibnamefont{and}
  \bibinfo{author}{\bibfnamefont{E.~S.} \bibnamefont{Polzik}}, in
  \emph{\bibinfo{booktitle}{Quantum Information with Continuous Variables}},
  edited by \bibinfo{editor}{\bibfnamefont{S.~L.} \bibnamefont{Braunstein}}
  \bibnamefont{and} \bibinfo{editor}{\bibfnamefont{A.~K.} \bibnamefont{Patel}}
  (\bibinfo{publisher}{Kluwer Academic Publishers}, \bibinfo{year}{2003}), pp.
  \bibinfo{pages}{231--265}.

\bibitem[{\citenamefont{Oblak et~al.}(2003)\citenamefont{Oblak, Mikkelsen,
  Tittel, Vershovski, S{\o}rensen, Petrov, \surname{Garrido Alzar}, and
  Polzik}}]{Oblak}
\bibinfo{author}{\bibfnamefont{D.}~\bibnamefont{Oblak}},
  \bibinfo{author}{\bibfnamefont{J.~K.} \bibnamefont{Mikkelsen}},
  \bibinfo{author}{\bibfnamefont{W.}~\bibnamefont{Tittel}},
  \bibinfo{author}{\bibfnamefont{A.~K.} \bibnamefont{Vershovski}},
  \bibinfo{author}{\bibfnamefont{J.~L.} \bibnamefont{S{\o}rensen}},
  \bibinfo{author}{\bibfnamefont{P.~G.} \bibnamefont{Petrov}},
  \bibinfo{author}{\bibfnamefont{C.~L.} \bibnamefont{\surname{Garrido Alzar}}},
  \bibnamefont{and} \bibinfo{author}{\bibfnamefont{E.~S.} \bibnamefont{Polzik}}
  (\bibinfo{year}{2003}), \eprint{quant-ph/0312165}.

\bibitem[{Foo({\natexlab{e}})}]{Footnote5}
\bibinfo{note}{Here we reintroduce implicitly the noise due to quantum
  fluctuations.}

\bibitem[{\citenamefont{de~Vries et~al.}(1998)\citenamefont{de~Vries, van
  Coevorden, and Lagendijk}}]{deVries}
\bibinfo{author}{\bibfnamefont{P.}~\bibnamefont{de~Vries}},
  \bibinfo{author}{\bibfnamefont{D.~V.} \bibnamefont{van Coevorden}},
  \bibnamefont{and}
  \bibinfo{author}{\bibfnamefont{A.}~\bibnamefont{Lagendijk}},
  \bibinfo{journal}{Rev. Mod. Phys.} \textbf{\bibinfo{volume}{70}},
  \bibinfo{pages}{447} (\bibinfo{year}{1998}).

\bibitem[{\citenamefont{Born and Wolf}(1999)}]{BornWolf}
\bibinfo{author}{\bibfnamefont{M.}~\bibnamefont{Born}} \bibnamefont{and}
  \bibinfo{author}{\bibfnamefont{E.}~\bibnamefont{Wolf}},
  \emph{\bibinfo{title}{Principles of Optics}} (\bibinfo{publisher}{Cambridge
  University Press}, \bibinfo{year}{1999}), chap. \bibinfo{chapter}{XIII},
  \bibinfo{edition}{seventh} ed.

\bibitem[{Foo({\natexlab{f}})}]{Footnote6}
\bibinfo{note}{For Cs atoms probed on the D$_{2}$ line the recoil time is $0.3
  ms$.}

\bibitem[{Foo({\natexlab{g}})}]{Footnote8}
\bibinfo{note}{Since we treat the scattered wave as a scalar the phase of the
  scattering amplitude has to be chosen with care.}

\bibitem[{\citenamefont{Mollow}(1975)}]{Mollow}
\bibinfo{author}{\bibfnamefont{B.~R.} \bibnamefont{Mollow}},
  \bibinfo{journal}{Phys. Rev. A} \textbf{\bibinfo{volume}{12}},
  \bibinfo{pages}{1919} (\bibinfo{year}{1975}).

\bibitem[{\citenamefont{Grimm et~al.}(2000)\citenamefont{Grimm, Weidem\"uller,
  and Ovchinnikov}}]{dipole}
\bibinfo{author}{\bibfnamefont{R.}~\bibnamefont{Grimm}},
  \bibinfo{author}{\bibfnamefont{M.}~\bibnamefont{Weidem\"uller}},
  \bibnamefont{and} \bibinfo{author}{\bibfnamefont{Y.~B.}
  \bibnamefont{Ovchinnikov}}, \bibinfo{journal}{Adv. in Atom., Mol. \& Opt.
  Phys.} \textbf{\bibinfo{volume}{42}}, \bibinfo{pages}{95}
  (\bibinfo{year}{2000}).

\bibitem[{Foo({\natexlab{h}})}]{Footnote10}
\bibinfo{note}{Here we neglect other potentials, like gravity, for simplicity.}

\bibitem[{Foo({\natexlab{i}})}]{Footnote11}
\bibinfo{note}{This corresponds to e.g. $8 \mu W$ probe power and a pulse
  duration of $2.9 \mu s$.}

\bibitem[{Foo({\natexlab{j}})}]{Footnote16}
\bibinfo{note}{Our simple analytical estimate at this length is already wrong
  by about a factor of 2}.

\bibitem[{\citenamefont{Smith et~al.}(2003)\citenamefont{Smith, Chaudhury, and
  Jessen}}]{Jessen}
\bibinfo{author}{\bibfnamefont{G.}~\bibnamefont{Smith}},
  \bibinfo{author}{\bibfnamefont{S.}~\bibnamefont{Chaudhury}},
  \bibnamefont{and} \bibinfo{author}{\bibfnamefont{P.~S.}
  \bibnamefont{Jessen}}, \bibinfo{journal}{J. Opt. B: Quantum Semiclassical
  Opt.} \textbf{\bibinfo{volume}{5}}, \bibinfo{pages}{323}
  (\bibinfo{year}{2003}).

\bibitem[{\citenamefont{Lye et~al.}(2003)\citenamefont{Lye, Hope, and
  Close}}]{Lye}
\bibinfo{author}{\bibfnamefont{J.~E.} \bibnamefont{Lye}},
  \bibinfo{author}{\bibfnamefont{J.~J.} \bibnamefont{Hope}}, \bibnamefont{and}
  \bibinfo{author}{\bibfnamefont{J.~D.} \bibnamefont{Close}},
  \bibinfo{journal}{Phys. Rev. A} \textbf{\bibinfo{volume}{67}},
  \bibinfo{pages}{043609} (\bibinfo{year}{2003}).

\bibitem[{Foo({\natexlab{k}})}]{Footnote12}
\bibinfo{note}{The scattering cross-section as an atomic property enters our
  calculation only once, consequently $\frac{\Gamma^{2}}{4\Delta^{2}}$ appears
  linearly here.}

\bibitem[{\citenamefont{Raymer and Mostowski}(1981)}]{Raymer}
\bibinfo{author}{\bibfnamefont{M.~G.} \bibnamefont{Raymer}} \bibnamefont{and}
  \bibinfo{author}{\bibfnamefont{J.}~\bibnamefont{Mostowski}},
  \bibinfo{journal}{Phys. Rev. A} \textbf{\bibinfo{volume}{24}},
  \bibinfo{pages}{1980} (\bibinfo{year}{1981}).

\bibitem[{Foo({\natexlab{l}})}]{Footnote13}
\bibinfo{note}{The density units of $(\lambda/2\pi)^3$ has to be smaller than
  $1$ for the model to be valid, otherwise the neglect of multiple scattering
  is not possible.}

\bibitem[{\citenamefont{Morice et~al.}(1995)\citenamefont{Morice, Castin, and
  Dalibard}}]{Morice}
\bibinfo{author}{\bibfnamefont{O.}~\bibnamefont{Morice}},
  \bibinfo{author}{\bibfnamefont{Y.}~\bibnamefont{Castin}}, \bibnamefont{and}
  \bibinfo{author}{\bibfnamefont{J.}~\bibnamefont{Dalibard}},
  \bibinfo{journal}{Phys. Rev. A} \textbf{\bibinfo{volume}{51}},
  \bibinfo{pages}{3896} (\bibinfo{year}{1995}).

\bibitem[{\citenamefont{Bouchoule and M{\o}lmer}(2002)}]{Bouchoule}
\bibinfo{author}{\bibfnamefont{I.}~\bibnamefont{Bouchoule}} \bibnamefont{and}
  \bibinfo{author}{\bibfnamefont{K.}~\bibnamefont{M{\o}lmer}},
  \bibinfo{journal}{Phys. Rev. A} \textbf{\bibinfo{volume}{66}},
  \bibinfo{pages}{043811} (\bibinfo{year}{2002}).

\bibitem[{\citenamefont{Rau et~al.}(2003)\citenamefont{Rau, Dunningham, and
  Burnett}}]{Burnett}
\bibinfo{author}{\bibfnamefont{A.~V.} \bibnamefont{Rau}},
  \bibinfo{author}{\bibfnamefont{J.~A.} \bibnamefont{Dunningham}},
  \bibnamefont{and} \bibinfo{author}{\bibfnamefont{K.}~\bibnamefont{Burnett}},
  \bibinfo{journal}{Science} \textbf{\bibinfo{volume}{301}},
  \bibinfo{pages}{1081} (\bibinfo{year}{2003}).

\bibitem[{Foo({\natexlab{m}})}]{Footnote14}
\bibinfo{note}{To lowest order in a multiple scattering expansion the dynamics
  of the density correlation function can be just described by recoil heating.}

\bibitem[{\citenamefont{Politzer}(1997)}]{Politzer}
\bibinfo{author}{\bibfnamefont{H.~D.} \bibnamefont{Politzer}},
  \bibinfo{journal}{Phys. Rev. A} \textbf{\bibinfo{volume}{55}},
  \bibinfo{pages}{1140} (\bibinfo{year}{1997}).

\bibitem[{\citenamefont{Inouye et~al.}(1999)\citenamefont{Inouye, Chikkatur,
  Stamper-Kurn, Stenger, Pritchard, and Ketterle}}]{Ketterle1}
\bibinfo{author}{\bibfnamefont{S.}~\bibnamefont{Inouye}},
  \bibinfo{author}{\bibfnamefont{A.}~\bibnamefont{Chikkatur}},
  \bibinfo{author}{\bibfnamefont{D.}~\bibnamefont{Stamper-Kurn}},
  \bibinfo{author}{\bibfnamefont{J.}~\bibnamefont{Stenger}},
  \bibinfo{author}{\bibfnamefont{D.}~\bibnamefont{Pritchard}},
  \bibnamefont{and} \bibinfo{author}{\bibfnamefont{W.}~\bibnamefont{Ketterle}},
  \bibinfo{journal}{Science} \textbf{\bibinfo{volume}{285}},
  \bibinfo{pages}{571} (\bibinfo{year}{1999}).

\bibitem[{\citenamefont{Schneble et~al.}(2003)\citenamefont{Schneble, Campbell,
  Streed, Boyd, Pritchard, and Ketterle}}]{Ketterle2}
\bibinfo{author}{\bibfnamefont{D.}~\bibnamefont{Schneble}},
  \bibinfo{author}{\bibfnamefont{G.~K.} \bibnamefont{Campbell}},
  \bibinfo{author}{\bibfnamefont{E.~W.} \bibnamefont{Streed}},
  \bibinfo{author}{\bibfnamefont{M.}~\bibnamefont{Boyd}},
  \bibinfo{author}{\bibfnamefont{D.~E.} \bibnamefont{Pritchard}},
  \bibnamefont{and} \bibinfo{author}{\bibfnamefont{W.}~\bibnamefont{Ketterle}}
  (\bibinfo{year}{2003}), \eprint{condmat/0311138}.

\bibitem[{\citenamefont{Horak et~al.}(1997)\citenamefont{Horak, Hechenblaikner,
  Gheri, Stecher, and Ritsch}}]{Horak}
\bibinfo{author}{\bibfnamefont{P.}~\bibnamefont{Horak}},
  \bibinfo{author}{\bibfnamefont{G.}~\bibnamefont{Hechenblaikner}},
  \bibinfo{author}{\bibfnamefont{K.~M.} \bibnamefont{Gheri}},
  \bibinfo{author}{\bibfnamefont{H.}~\bibnamefont{Stecher}}, \bibnamefont{and}
  \bibinfo{author}{\bibfnamefont{H.}~\bibnamefont{Ritsch}},
  \bibinfo{journal}{Phys. Rev. Lett.} \textbf{\bibinfo{volume}{79}},
  \bibinfo{pages}{4974} (\bibinfo{year}{1997}).

\bibitem[{\citenamefont{Vuletic and Chu}(2000)}]{Vuletic}
\bibinfo{author}{\bibfnamefont{V.}~\bibnamefont{Vuletic}} \bibnamefont{and}
  \bibinfo{author}{\bibfnamefont{S.}~\bibnamefont{Chu}},
  \bibinfo{journal}{Phys. Rev. Lett.} \textbf{\bibinfo{volume}{84}},
  \bibinfo{pages}{3787} (\bibinfo{year}{2000}).

\bibitem[{\citenamefont{Nagorny et~al.}(2003)\citenamefont{Nagorny,
  Els\"{a}sser, and Hemmerich}}]{Hemmerich}
\bibinfo{author}{\bibfnamefont{B.}~\bibnamefont{Nagorny}},
  \bibinfo{author}{\bibfnamefont{T.}~\bibnamefont{Els\"{a}sser}},
  \bibnamefont{and}
  \bibinfo{author}{\bibfnamefont{A.}~\bibnamefont{Hemmerich}},
  \bibinfo{journal}{Phys. Rev. Lett.} \textbf{\bibinfo{volume}{91}},
  \bibinfo{pages}{153003} (\bibinfo{year}{2003}).

\end{thebibliography}
\end{document}